\documentclass[twocolumn, times]{aastex631}

\usepackage{threeparttable} 
\usepackage{float}
\usepackage{amsmath}
\usepackage{rotating}
\usepackage{xcolor}
\usepackage{amssymb}
\usepackage{colortbl}

\usepackage{ulem}
\usepackage{array, boldline, makecell, booktabs}

\renewcommand{\textbf}[1]{#1}

\begin{document}


\title{Quantified Estimation of Molecular Detections across Different Classes of Neptunian Atmospheres Using Cross-Correlation Spectroscopy: Prospects for Future Extremely Large Telescopes with High-Resolution Spectrographs}

\correspondingauthor{Liton Majumdar}
\email{liton@niser.ac.in, dr.liton.majumdar@gmail.com}

\author[0000-0002-7033-209X]{Dwaipayan Dubey}
\affiliation{Universitäts-Sternwarte, Fakultät für Physik, Ludwig-Maximilians-Universität München, Scheinerstr. 1, D-81679 München, Germany}
\affiliation{Exzellenzcluster `Origins’, Boltzmannstr. 2, D-85748 Garching, Germany}
\affiliation{Exoplanets and Planetary Formation Group, School of Earth and Planetary Sciences, National Institute of Science Education and Research, Jatni 752050, Odisha, India}
\affiliation{Homi Bhabha National Institute, Training School Complex, Anushaktinagar, Mumbai 400094, India}

\author[0000-0001-7031-8039]{Liton Majumdar}
\affiliation{Exoplanets and Planetary Formation Group, School of Earth and Planetary Sciences, National Institute of Science Education and Research, Jatni 752050, Odisha, India}
\affiliation{Homi Bhabha National Institute, Training School Complex, Anushaktinagar, Mumbai 400094, India}

\author[0000-0002-5627-5471]{Charles Beichman}
\affiliation{IPAC, California Institute of Technology, 1200 E California Boulevard, Pasadena, CA 91125, USA} 

\author[0000-0003-0787-1610]{Geoffrey A. Blake}
\affiliation{Division of Geological \& Planetary Sciences, California Institute of Technology, Pasadena, CA 91125, USA}

\author[0000-0002-1871-6264]{Gautam Vasisht}
\affiliation{Jet Propulsion Laboratory, California Institute of Technology, 4800 Oak Grove Dr, Pasadena, CA 91109, USA}

\author[0000-0002-1493-300X]{Thomas Henning}
\affiliation{Max Planck Institute for Astronomy, Königstuhl 17, D-69117 Heidelberg, Germany}

\begin{abstract}

Neptune-size exoplanets are less studied as characterizing their atmospheres presents challenges due to their relatively small radius and atmospheric scale height. As the most common outcome of planet formation, these planets are crucial for understanding planetary formation, migration theories, atmospheric composition, and potential habitability. Their diverse atmospheres, influenced by equilibrium temperature, composition, and cloud presence, offer unique opportunities to study atmospheric dynamics and chemistry. While low-resolution spectroscopy struggles with atmospheric characterization due to clouds, high-resolution observations provide detailed analysis of the atmospheres by detecting molecular lines beyond the cloud deck. This study investigates four subclasses of Neptune atmospheres: HAT-P-11 b (warm Neptune), HD 63433 c (warm sub-Neptune), K2-25 b (temperate Neptune), and TOI-270 d (temperate sub-Neptune), using six ground-based spectrographs: GIANO-B, CARMENES, IGRINS, HISPEC, MODHIS, and ANDES over one and three transits. Our simulation integrates the chemical kinetics model,  VULCAN with the 1-D line-by-line radiative transfer model, petitRADTRANS, and estimates detection significance using the ground-based noise simulator, SPECTR. We aim to predict how future  Extremely Large Telescopes (ELTs) such as TMT (MODHIS) and E-ELT (ANDES) can utilize their higher resolving powers and larger collecting areas to surpass current observatories in detecting molecular bands. We highlight the importance of photochemistry in these atmospheres and demonstrate how ELTs will help further in constraining nitrogen and sulfur chemistry. Finally, we present a comprehensive picture of cloud presence in the atmospheres and its impact on molecular detectability in Neptune-class atmospheres.

\end{abstract}


\section{Introduction} \label{sec:intro}


With the detection of over 5,600 planets to date, the primary focus in exoplanetary science has shifted from planet detection to the characterization of planetary atmospheres. Understanding these atmospheres has become crucial in linking planet formation history to atmospheric dynamics and evolution (see \cite{oberg2011effects,mordasini2016imprint,espinoza2017metal,madhusudhan2017atmospheric,Dash2022,reggiani2022evidence}). To this aim, spectroscopic observations have made substantial progress over the past few years, making it one of the most exciting avenues in the field \citep{crossfield2015observations}. Transit spectroscopy has proven to be the primary and most favored method for atmospheric characterization. In addition to the myriad of observations with past and current state-of-the-art space-borne facilities, high-resolution spectroscopy (HRS) from ground-based observatories has become important due to its ability to detect molecular bands individually and provides additional benefits through the in-depth analysis of atmospheric chemistry \citep{snellen2010orbital, wyttenbach2015spectrally, brogi2016rotation, brogi2018exoplanet, alonso2019multiple,sanchez2019water, giacobbe2021five}.

HRS methods have advanced significantly over the past decade, becoming essential for studying planetary atmospheres in optical and near-infrared bands. HRS stands out as an interesting approach, making it possible to identify specific line features within the broadband molecular bands \citep{2024ApJ...972..165D}. Nevertheless, the effectiveness of observations is affected by the significant challenge of reducing the impact of strong telluric and stellar backgrounds on the planetary signal \citep{sanchez2019water}. \textbf{To address this issue, the Doppler shift of the planetary spectrum is induced by the planet's reflex orbital motion around its host star to mitigate the interference from these background signals.}
Ultimately, the detection significance of molecular bands is computed using the cross-correlation technique where the observed spectra are cross-correlated with the template spectra of different molecules \citep{snellen2010orbital, brogi2012signature, birkby2013monthly}.

\begin{figure}
\centering
	\includegraphics[width=\columnwidth]{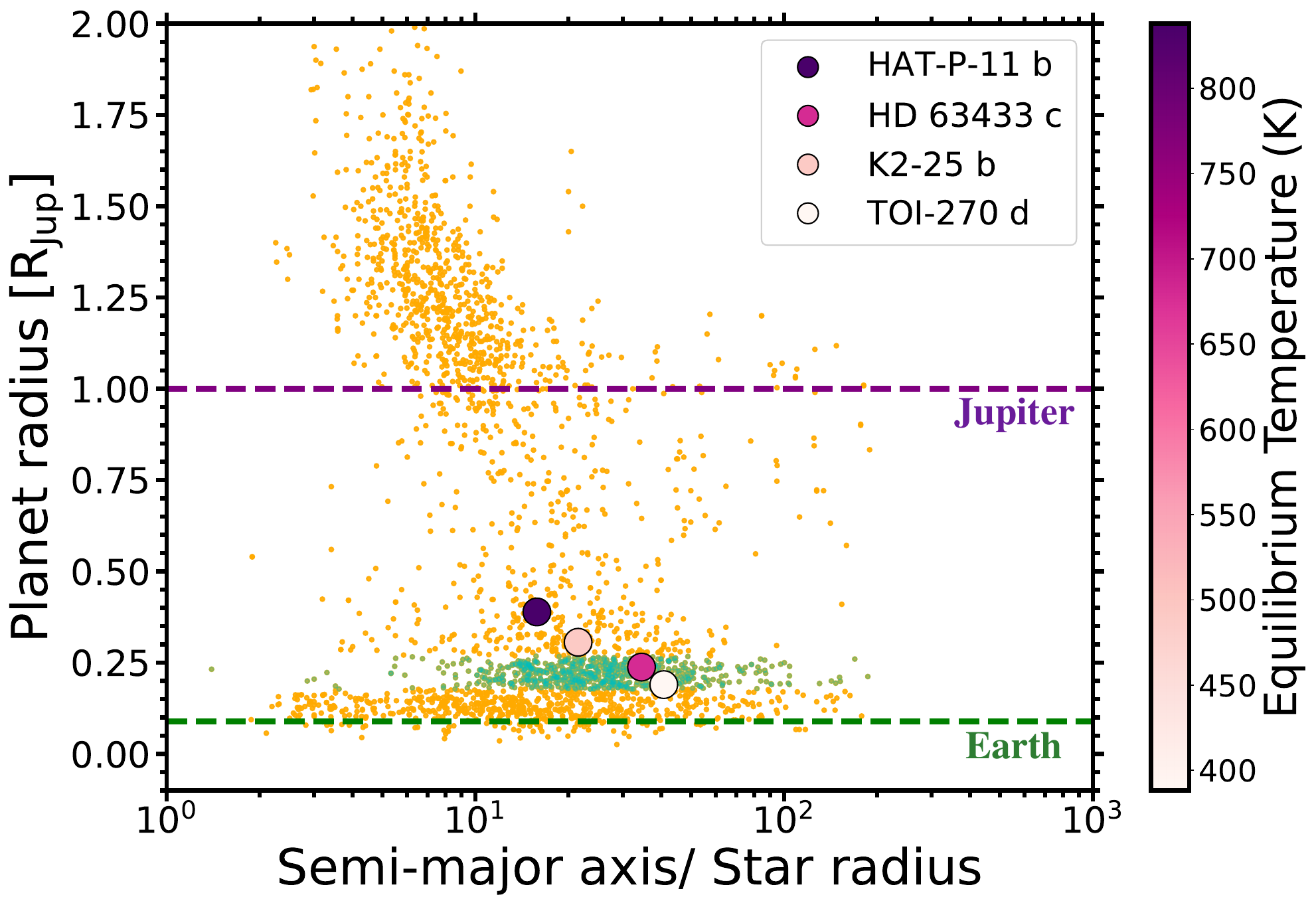}
  \caption{Planet radius as a function of semi-major axis/host star radius. The `yellow' scatter points indicate the population of exoplanets detected to date, with data sourced from the \href{https://exoplanetarchive.ipac.caltech.edu/}{NASA Exoplanet Archive} on June 23, 2024. The `cyan' points highlight planets with radii between 2-3 $\mathrm{R_{\oplus}}$, generally categorized as Neptune-class planets. For reference, the radii of Earth and Jupiter are included to illustrate the location of the selected planets within the entire exoplanet population.}
   \label{fig:population}
\end{figure}

The advancement in HRS started with the first detection of Na in the atmospheres of HD 189733 b \citep{redfield2008sodium} and HD 209458 b \citep{snellen2008ground} with numerous subsequent observations aimed at molecular detection \citep{snellen2010orbital, brogi2012signature, birkby2013detection, hawker2018evidence, alonso2019multiple, guilluy2019exoplanet, cabot2019robustness, giacobbe2021five, carleo2022gaps}. Later on, the emphasis has progressed from individual species detection \citep{snellen2010orbital,brogi2012signature,birkby2013detection,de2013detection,lockwood2014near} to simultaneously detecting multiple species \citep{hawker2018evidence,cabot2019robustness,giacobbe2021five}. So far, the most accurate constraints have been derived from investigations focused on hot and ultra-hot Jupiters, benefiting from their expanded atmospheres that facilitate the applicability of transmission spectroscopy. The most common planet populations have radii in the range of 1-4 $\mathrm{R_{\oplus}}$, covering two distinct classes of planets separated by Fulton gap: Neptunes (2-3 $\mathrm{R_{\oplus}}$) with $\mathrm{H_2}$/$\mathrm{He}$-rich gas envelopes, and super-Earths (up to 1.5 $\mathrm{R_{\oplus}}$) with rocky cores or magma oceans \citep{fulton2017california,owen2017evaporation,ginzburg2018core}. Despite being the most common population with atmospheres, efforts to determine the atmospheric composition of Neptunian planets have presented considerable challenges. This difficulty arises from the compositional degeneracy between different Neptunian subclasses, $\mathrm{H_2}$-rich water worlds and iron-rich thin atmosphere worlds, making it hard to differentiate between them \citep{adams2008ocean,madhusudhan2021habitability,kempton2023water}. A proposed solution to address this issue involves the precise atmospheric characterization of such objects, which could potentially resolve the problem of compositional ambiguity \citep{adams2008ocean}. These cooler planets are anticipated to have clouds in their atmospheres that can obscure molecular features, posing challenges for low-resolution spectroscopy \citep{knutson2014featureless,kreidberg2014clouds,crossfield2017trends}. Conversely, HRS can probe molecular line cores extending beyond the cloud deck, offering a more reliable method to trace molecular signatures in Neptunian atmospheres \citep{de2014identifying,kempton2014high,gandhi2020seeing,hood2020prospects}. Therefore, the question of the hour revolves around whether high-resolution observations can excel in our understanding of the atmospheres of such planets.  Being in the nexus of terrestrial planets and big gas giants, Neptune class planets pave our understanding in tracing planet formation mechanisms and atmospheric evolution history. They represent a crucial transitional regime that bridges the gap between these different planetary types. This transitional nature not only enhances our understanding of planetary atmospheres but also sheds light on the complex processes governing planetary formation and migration within diverse exoplanetary architectures. The Neptune class covers a diverse range of systems, spanning from warm to temperate atmospheres. Consequently, it is considered the optimal selection for thorough studies aimed at gaining a detailed understanding of planetary systems as a whole.

\begin{table*}
\centering
\caption{Star and planetary system parameters used in this study and their values}
\begin{tabular*}{2\columnwidth}{@{\extracolsep{\fill}}lcccccc@{}}

\hline
\hline
\noalign{\smallskip}
&\multicolumn{4}{c}{\textbf{\hspace{2.0cm}STAR PARAMETERS}}  & \\
\noalign{\smallskip}
\hline
\noalign{\smallskip}
& \multicolumn{2}{c}{HAT-P-11} & \multicolumn{3}{c}{\hspace{1cm}HD 63433} \\
\noalign{\smallskip}
\hlineB{3.5}
\noalign{\smallskip}
\multicolumn{1}{l|}{Parameters} & Values & \hspace{1cm}References & \multicolumn{1}{|c}{} & \hspace{-0.5cm}Values & \hspace{1cm}References \\
\noalign{\smallskip}
 \hline
\noalign{\smallskip}
    \multicolumn{1}{l|}{$M_*$/$M_{\odot}$} & 0.809 & \hspace{1cm} \cite{2020MNRAS.497.2096X}& \multicolumn{1}{|c}{} & \hspace{-0.5cm}0.99 & \hspace{1cm}\cite{2024AJ....167...54C}\\
\multicolumn{1}{l|}{$R_*$/$R_{\odot}$} & 0.68 & \hspace{1cm}\cite{2018AJ....155..255Y} & \multicolumn{1}{|c}{} & \hspace{-0.5cm}0.91 & \hspace{1cm}\cite{2024AJ....167...54C}\\
\multicolumn{1}{l|}{$T_*$ [K]} & 4780  & \hspace{1cm}\cite{2018AJ....155..255Y} & \multicolumn{1}{|c}{} & \hspace{-0.5cm}5640  & \hspace{1cm}\cite{2020AJ....160..179M}\\
\multicolumn{1}{l|}{Rotation period [days]} & 29.2 & \hspace{1cm}\cite{2018AJ....155..255Y} & \multicolumn{1}{|c}{} & \hspace{-0.5cm}6.4 & \hspace{1cm}\cite{2024AJ....167...54C}\\
\multicolumn{1}{l|}{Radial velocity of system [km/s]} & -63.24 & \hspace{1cm}\cite{brown2018gaia} & \multicolumn{1}{|c}{} & \hspace{-0.5cm}-16.07 & \hspace{1cm}\cite{2024AJ....167...54C}\\

\noalign{\smallskip}
\hline
\hline
\noalign{\smallskip}
\noalign{\smallskip}
& \multicolumn{2}{c}{K2-25} & \multicolumn{3}{c}{\hspace{1cm}TOI-270} \\
\noalign{\smallskip}
\hlineB{3.5}
\noalign{\smallskip}
\multicolumn{1}{l|}{Parameters} & Values & \hspace{1cm}References & \multicolumn{1}{|c}{} & \hspace{-0.5cm}Values & \hspace{1cm}References \\
\noalign{\smallskip}
 \hline
\noalign{\smallskip}
    \multicolumn{1}{l|}{$M_*$/$M_{\odot}$} & 0.26 & \hspace{1cm}\cite{2020AJ....160..192S} & \multicolumn{1}{|c}{} & \hspace{-0.5cm}0.39 & \hspace{1cm}\cite{2021MNRAS.507.2154V}\\
\multicolumn{1}{l|}{$R_*$/$R_{\odot}$} & 0.29 & \hspace{1cm}\cite{2020AJ....160..192S} & \multicolumn{1}{|c}{} & \hspace{-0.5cm}0.38 & \hspace{1cm}\cite{2023AJ....165...84M}\\
\multicolumn{1}{l|}{$T_*$ [K]} & 3207  & \hspace{1cm}\cite{2020AJ....160..192S} & \multicolumn{1}{|c}{} & \hspace{-0.5cm}3506  & \hspace{1cm}\cite{2021MNRAS.507.2154V}\\
\multicolumn{1}{l|}{Rotation period [days]} & 1.878 & \hspace{1cm}\cite{2020AJ....160..192S} & \multicolumn{1}{|c}{} & \hspace{-0.5cm}--  & \hspace{1cm}-- \\
\multicolumn{1}{l|}{Radial velocity of system [km/s]} & 38.64 & \hspace{1cm}\cite{2016ApJ...818...46M} & \multicolumn{1}{|c}{} & \hspace{-0.5cm}25.90 & \hspace{1cm}\cite{vallenari2023gaia}\\
\noalign{\smallskip}
\hline
\hline
\noalign{\smallskip}\noalign{\smallskip}\noalign{\smallskip}
& \multicolumn{4}{c}{\textbf{\hspace{2.0cm}PLANET PARAMETERS}} &  \\
\noalign{\smallskip}
\hline
\noalign{\smallskip}
& \multicolumn{2}{c}{Warm Neptune: HAT-P-11 b} & \multicolumn{3}{c}{\hspace{0.2cm}Warm sub-Neptune: HD 63433 c} \\
\noalign{\smallskip}
\hlineB{3.5}
\noalign{\smallskip}
\multicolumn{1}{l|}{$M\mathrm{_p}$/$M\mathrm{_{Jup}}$} & 0.09 & \hspace{1cm}\cite{2017AJ....153..136S} & \multicolumn{1}{|c}{} & \hspace{-0.5cm}0.0239 & \hspace{1cm}\cite{2020AJ....160..179M}\\
\multicolumn{1}{l|}{$R\mathrm{_p}$/$R\mathrm{_{Jup}}$} & 0.389 & \hspace{1cm}\cite{2018AJ....155..255Y} & \multicolumn{1}{|c}{} & \hspace{-0.5cm}0.238 & \hspace{1cm}\cite{2020AJ....160..179M}\\
\multicolumn{1}{l|}{$T\mathrm{_{eq}}$ [K]} & 838 & \hspace{1cm}\cite{2011MNRAS.417.2166S} & \multicolumn{1}{|c}{} & \hspace{-0.5cm}621.93 & \hspace{1cm} \href{https://exofop.ipac.caltech.edu/tess/view\_toi.php}{ExoFOP-TESS TOI}\\
\multicolumn{1}{l|}{C/O} & 0.97 & \hspace{1cm}\cite{chachan2019hubble} & \multicolumn{1}{|c}{} & \hspace{-0.5cm} 0.55 & \hspace{1cm} --\\
\multicolumn{1}{l|}{[Fe/H]} & -0.98 & \hspace{1cm}\cite{chachan2019hubble} & \multicolumn{1}{|c}{} & \hspace{-0.5cm} 0.05 & \hspace{1cm} \cite{dai2020tess}\\
\multicolumn{1}{l|}{$a$ [au]} & 0.053 & \hspace{1cm}\cite{2011MNRAS.417.2166S} & \multicolumn{1}{|c}{} & \hspace{-0.5cm}0.145 & \hspace{1cm}\cite{2024AJ....167...54C}\\
\multicolumn{1}{l|}{Period [days]} & 4.89 & \hspace{1cm}\cite{2018AJ....155..255Y} & \multicolumn{1}{|c}{} & \hspace{-0.5cm}20.54 & \hspace{1cm}\cite{2024AJ....167...54C}\\
\multicolumn{1}{l|}{Distance [pc]} & 37.76 & \hspace{1cm}\cite{2007ASSL..350.....V} & \multicolumn{1}{|c}{} & \hspace{-0.5cm}22.4 & \hspace{1cm}\cite{2024AJ....167...54C}\\
\multicolumn{1}{l|}{Transit Duration [hours]} & 2.364 & \hspace{1cm}\cite{2016ApJ...822...86M} & \multicolumn{1}{|c}{} & \hspace{-0.5cm}4.10 & \hspace{1cm}\cite{2024AJ....167...54C}\\
\noalign{\smallskip}
\hline
\hline
\noalign{\smallskip}
\noalign{\smallskip}
& \multicolumn{2}{c}{Temperate Neptune: K2-25 b} & \multicolumn{3}{c}{\hspace{0.2cm}Temperate sub-Neptune: TOI-270 d} \\
\noalign{\smallskip}
\hlineB{3.5}
\noalign{\smallskip}
\multicolumn{1}{l|}{$M\mathrm{_p}$/$M\mathrm{_{Jup}}$} & 0.0771 & \hspace{1cm}\cite{2020AJ....160..192S} & \multicolumn{1}{|c}{} & \hspace{-0.5cm}0.015 & \hspace{1cm}\cite{2021MNRAS.507.2154V}\\
\multicolumn{1}{l|}{$R\mathrm{_p}$/$R\mathrm{_{Jup}}$} & 0.306 & \hspace{1cm}\cite{2020AJ....160..192S} & \multicolumn{1}{|c}{} & \hspace{-0.5cm}0.19 & \hspace{1cm}\cite{2021MNRAS.507.2154V}\\
\multicolumn{1}{l|}{$T\mathrm{_{eq}}$ [K]} & 494 & \hspace{1cm}\cite{2020AJ....160..192S} & \multicolumn{1}{|c}{} & \hspace{-0.5cm}387 & \hspace{1cm}\cite{2021MNRAS.507.2154V}\\
\multicolumn{1}{l|}{C/O} & 0.22 & \hspace{1cm}\cite{blain20211d} & \multicolumn{1}{|c}{} & \hspace{-0.5cm} 0.47 & \hspace{1cm} \cite{benneke2024jwst}\\
\multicolumn{1}{l|}{[Fe/H]} & 0.15 & \hspace{1cm}\cite{stefansson2020habitable} & \multicolumn{1}{|c}{} & \hspace{-0.5cm} 2.35 & \hspace{1cm} \cite{benneke2024jwst}\\
\multicolumn{1}{l|}{$a$ [au]} & 0.029 & \hspace{1cm}\cite{2020AJ....160..192S} & \multicolumn{1}{|c}{} & \hspace{-0.5cm}0.072 & \hspace{1cm}\cite{2021MNRAS.507.2154V}\\
\multicolumn{1}{l|}{Period [days]} & 3.48 & \hspace{1cm}\cite{2020AJ....160..192S} & \multicolumn{1}{|c}{} & \hspace{-0.5cm}11.38 & \hspace{1cm}\cite{2021MNRAS.507.2154V}\\
\multicolumn{1}{l|}{Distance [pc]} & 44.96 & \hspace{1cm}\cite{2016ApJ...818...46M} & \multicolumn{1}{|c}{} & \hspace{-0.5cm}22.48 & \hspace{1cm}\cite{2019NatAs...3.1099G}\\
\multicolumn{1}{l|}{Transit Duration [hours]} & 0.7637 & \hspace{1cm}\cite{2020AJ....160..192S} & \multicolumn{1}{|c}{} & \hspace{-0.5cm}2.148 & \hspace{1cm}\cite{2019NatAs...3.1099G}\\
\noalign{\smallskip}
\hline
\hline
\label{tab:system}
\end{tabular*}
\end{table*}


HAT-P-11 b, HD 63433 c, K2-25 b, and TOI-270 d are four Neptune candidates of our interest due to their variable physical structures, equilibrium temperatures ($T_{\mathrm{eq}}$), and host star properties. Based on the $T_{\mathrm{eq}}$ and size of the planet, we classify the four planets into four subcategories: warm Neptune (HAT-P-11 b), warm sub-Neptune (HD 63433 c), temperate Neptune (K2-25 b), and temperate sub-Neptune (TOI-270 d). See Figure \ref{fig:population} for an overview of the systems. The presence of a hot extended atmosphere makes HAT-P-11 b a well-suited candidate for transmission spectroscopy. On top of that, the $T_{\mathrm{eq}}$ of this planet is conducive to the absence of a cloudy atmosphere, thereby allowing for more feature-rich transmission spectra \citep{gao2021aerosols}. On the other hand, HD 63433 c, K2-25 b, and TOI-270 d possess a temperature that is suitable for the photochemically produced hydrocarbon hazes that can mute molecular features. HAT-P-11 b and TOI-270 d were earlier detected with $\mathrm{H_2O}$ features from space observations with HST and \textit{Spitzer} \citep{fraine2014water,mikal2023hubble}. Recent observations by JWST have revealed the presence of $\mathrm{CH_4}$, $\mathrm{H_2O}$, $\mathrm{CO_2}$, and a tentative signature of $\mathrm{CS_2}$ in the metal-rich atmosphere of TOI-270 d \citep{benneke2024jwst}. However, no attempt has been made yet to robustly deduce molecular fingerprints in their atmospheres using high-resolution observation except for HAT-P-11 b. The GAPS program at TNG has identified features of $\mathrm{H_2O}$ and $\mathrm{NH_3}$ on HAT-P-11 b, along with minimal detections of $\mathrm{CH_4}$ and $\mathrm{CO_2}$ \citep{basilicata2024gaps}. Further exploration of gas giant atmospheres with TNG has provided evidence of leading disequilibrium chemistry in their atmospheres \citep{carleo2022gaps,guilluy2022gaps}. So far, HD 63433 c and K2-25 b were primarily focused on studying atmospheric escape \citep{rockcliffe2021lyalpha,zhang2022detection,orell2024mopys,masson2024probing,rockcliffe2021lyalpha} and have not undergone detailed atmospheric characterization. 

Considering the diversity of Neptunian planets in mind and focusing on HRS alongside the state-of-the-art JWST observations, we aim to explore the different classes of Neptunian atmospheres. We investigate how current and forthcoming ground-based observations can advance in detecting various molecules within their atmospheres and enhance our understanding of the underlying chemical processes. In this study, we present the diverse atmospheric chemistry of HAT-P-11 b, HD 63433 c, K2-25 b, and TOI-270 d, along with a complementary analysis of molecular detectability utilizing six high-resolution ground-based spectrographs: GIANO-B (Telescopio Nazionale Galileo: TNG)\footnote{\href{https://www.tng.iac.es/instruments/giano-b/}{https://www.tng.iac.es/instruments/giano-b/}} \citep{oliva2012giano}, CARMENES (Centro Astronomico Hispano Alemán: CAHA)\footnote{\href{https://www.caha.es/telescope-3-5m/carmenes}{https://www.caha.es/telescope-3-5m/carmenes}} \citep{quirrenbach2014carmenes,quirrenbach2018carmenes}, IGRINS (The International Gemini Observatory)\footnote{\href{https://www.gemini.edu/instrumentation/igrins}{https://www.gemini.edu/instrumentation/igrins}} \citep{yuk2010preliminary}, HISPEC (Keck)\footnote{\href{https://www.ucobservatories.org/projects\_at/keck-hispec/}{https://www.ucobservatories.org/projects\_at/keck-hispec/}} \citep{Mawet2019High}, MODHIS (Thirty Meter Telescope: TMT)\footnote{\href{https://www.tmt.org/page/modhis}{https://www.tmt.org/page/modhis}} \citep{Mawet2019High}, and ANDES (European Extremely Large Telescope: E-ELT)\footnote{\href{https://elt.eso.org/instrument/ANDES/}{https://elt.eso.org/instrument/ANDES/}} \citep{marconi2022andes}. We outline our methods in Section \ref{sec:methods}, detailing the simulation and evolution of planetary atmospheres using the 1-D self-consistent forward model, petitCODE \citep{molliere2015model,molliere2017modeling}, and the 1-D chemical kinetics model, VULCAN \citep{Tsai_2017,Tsai_2021}. This section explains our assumptions for estimating molecular detection significance through synthetic observations, utilizing a high-resolution radiative transfer method using the petitRADTRANS code \citep{refId0, molliere2020}, and a noise simulator model for ground-based observatories, \text{using} the Spectral Planetary ELT Calculator (SPECTR) package \citep{currie2023there}. A brief overview of high-resolution opacity calculation from molecular line lists has also been discussed here. Section \ref{sec:results} presents the potential detection of various molecular bands across different atmospheres for different numbers of transits. Finally, we summarize our findings in Section \ref{sec:conclusion}.

\section{Methods}
\label{sec:methods}

This section presents our methodology for characterizing various classes of Neptunian atmospheres using cross-correlation spectroscopy. Our approach begins with self-consistent calculations of atmospheric structures and chemistry for all planets, followed by a summary of the pipeline used for molecular detection in their atmospheres. Detailed descriptions are provided in the following subsections.

\begin{figure}
\centering
	\includegraphics[width=\columnwidth]{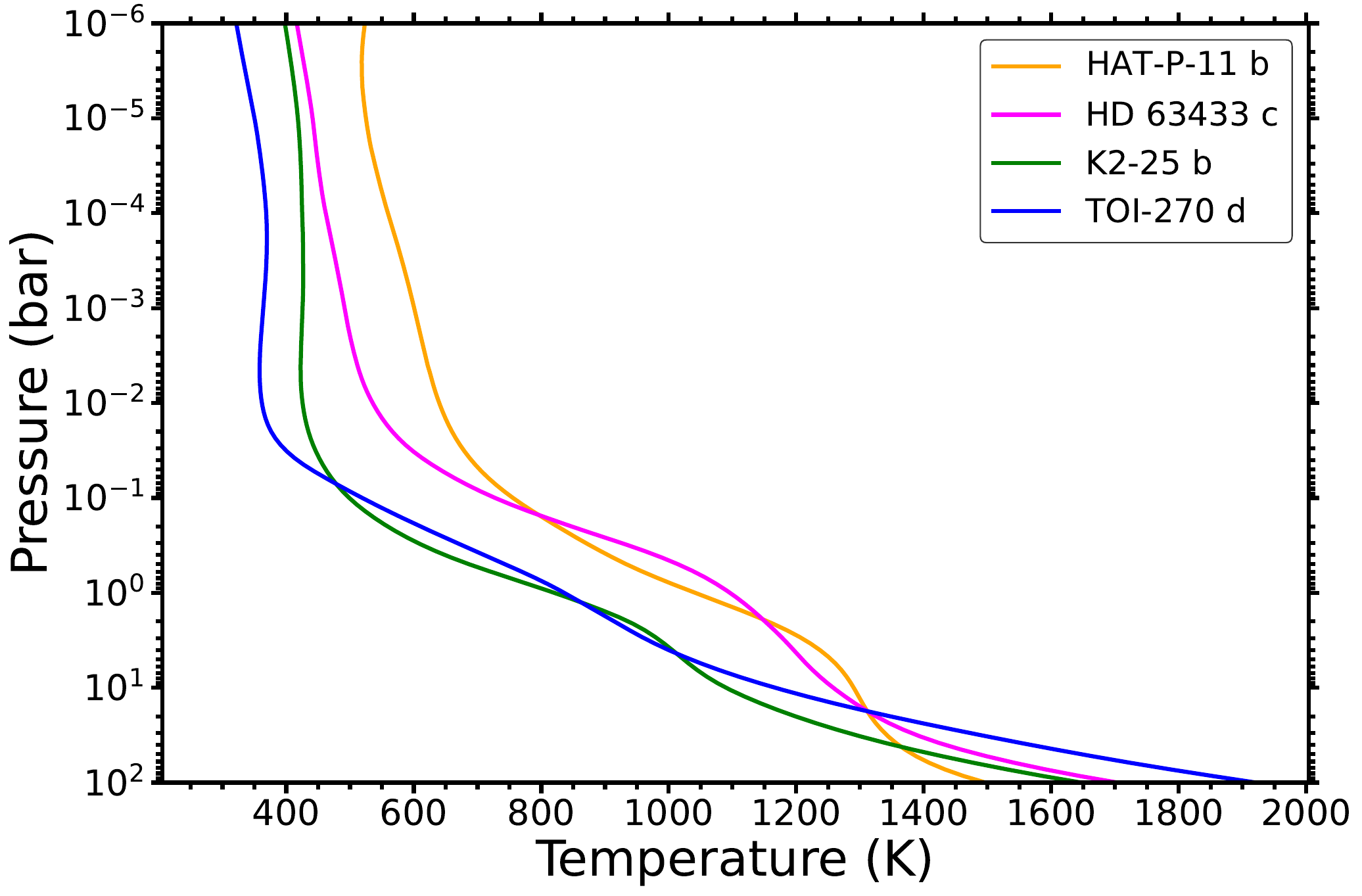}
  \caption{Self-consistent pressure-temperature (PT) structures for different Netune class planets. It is constructed using the 1-D radiative-convective equilibrium model, petitCODE. This simulation utilized the system parameters listed in Table \ref{tab:system} and elemental abundances specified in Section \ref{sec:atmosphere}. In the legend, the planets are ordered sequentially from highest (HAT-P-11 b) to lowest (TOI-270 d) planetary $T_{\mathrm{eq}}$. All the planets exhibit non-inverted PT structures.}
   \label{fig:TP_profile}
\end{figure}

\begin{figure*}
\centering
	\includegraphics[width=1.75\columnwidth]{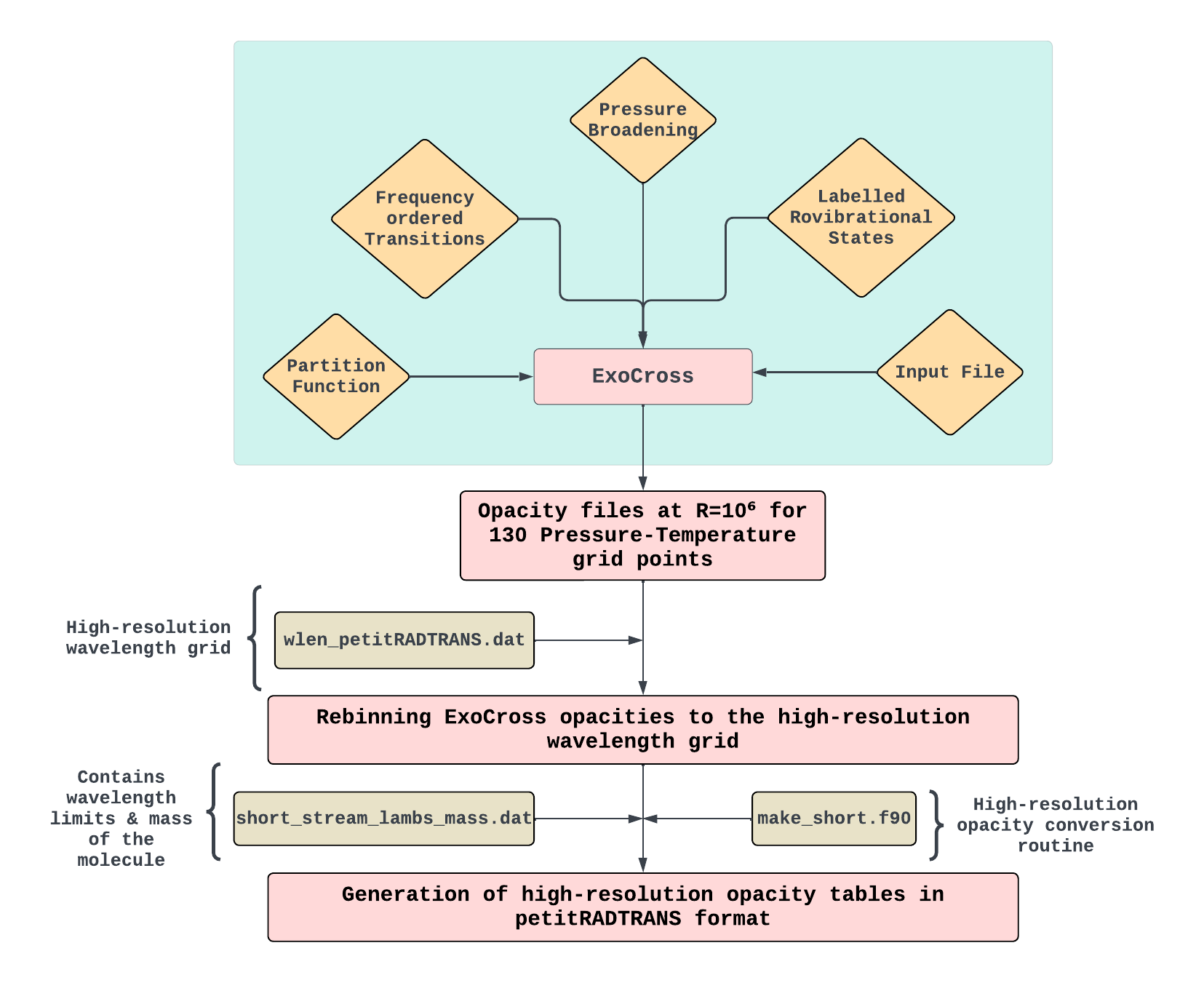}
  \caption{The specific routine followed to calculate high-resolution line-by-line (R = $\mathrm{10^6}$) opacity for certain molecules, as listed in Table \ref{tab:opacity}. ExoMol or HITRAN line lists are used to generate the molecular opacity in petitRADTRANS format for 130 pressure-temperature grid points. Molecule-specific broadening parameters are also being employed.}
   \label{fig:flowchart}
\end{figure*}

\subsection{Atmospheric structure and leading chemistry}
\label{sec:atmosphere}

We self-consistently generated the pressure-temperature (PT) structures for all planets using the 1-D forward model petitCODE (see Figure \ref{fig:TP_profile}). petitCODE considers two simplistic assumptions to simulate the planetary atmospheres accurately: (1) radiative-convective equilibrium and (2) thermochemical equilibrium. The stellar and planetary system parameters are pivotal for simulating the atmospheric pressure-temperature (PT) structures and calculating the equilibrium mixing ratio profiles (VMR) of molecules. The system parameters are listed in Table \ref{tab:system}. The atmospheric carbon-to-oxygen (C/O) ratio and metallicity ([Fe/H]) of planets are fundamental parameters for understanding and modeling the composition and thermal structure of planetary atmospheres. They determine the relative abundances of various molecules, thereby influencing the PT structures of their atmospheres. For our study, we incorporated the most recent measurements of C/O ratios and [Fe/H] values specific to HAT-P-11 b \citep{chachan2019hubble} and TOI-270 d \citep{benneke2024jwst}. For HD 63433 c, we adopted the solar C/O ratio. On the other hand, for K2-25 b, we pursued the same C/O ratio measured on K2-18 b \citep{blain20211d}. This assumption is to leverage existing knowledge and similarities between the target planets and their respective stellar systems to infer their atmospheric properties accurately. Regarding [Fe/H] of these two planets, we used their respective stellar metallicity values, which are close to solar ([Fe/H] = 0.05 for HD 63433 c \citep{dai2020tess} and 0.15 for K2-25 b \citep{stefansson2020habitable}. Here, [Fe/H] stands for the metallicity in the log scale where [Fe/H] = 0 represents the solar value. We utilized petitCODE with its default chemical input parameters for our calculations with $\mathrm{H_2}$-$\mathrm{He}$ dominated atmosphere, incorporating opacity sources such as \textbf{$\mathrm{CH_4}$, $\mathrm{H_2O}$, $\mathrm{CO_2}$, $\mathrm{HCN}$, CO, $\mathrm{H_2S}$, $\mathrm{NH_3}$, $\mathrm{C_2H_2}$ (HITRAN; see \cite{HITRAN2020}), Na, K (VALD3; see \cite{piskunov1995vald}), TiO (ExoMol; see \cite{mckemmish2019exomol}), and VO (ExoMol; see \cite{bowesman2024exomol})} in the planetary atmospheres to perform the radiative transfer calculations. Additionally, collision-induced absorption (CIA) from $\mathrm{H_2}$-$\mathrm{H_2}$ and $\mathrm{H_2}$-$\mathrm{He}$ was considered. For determining the C/O ratio, we adopted a methodology consistent with \cite{madhusudhan2012c}, \cite{molliere2015model}, \cite{woitke2018equilibrium}, \cite{molaverdikhani2019cold}, \cite{dubey2023polycyclic}, and \cite{2024ApJ...972..165D}, adjusting the oxygen elemental abundance while keeping the carbon elemental abundance constant (see the C/O ratios from Table \ref{tab:system}). 


\textbf{We integrated the PT structures generated by petitCODE (see Figure \ref{fig:TP_profile}) into the 1-D chemical kinetics model, VULCAN\footnote{\href{https://github.com/exoclime/VULCAN}{https://github.com/exoclime/VULCAN}}, to simulate the evolution of atmospheric disequilibrium chemistry. VULCAN begins with equilibrium chemistry profiles generated using Fastchem\footnote{\href{https://github.com/exoclime/FastChem}{https://github.com/exoclime/FastChem}} \citep{stock2018fastchem}, based on the elemental abundances of the respective planets and PT structures generated from petitCODE, and allows the atmosphere to evolve kinetically until convergence is achieved. The model accounts for different physical processes such as molecular diffusion, atmospheric mixing (by implementing the Eddy diffusion coefficient, $K_{\mathrm{zz}}$), and photochemistry.} The impact of stellar UV flux is crucial for tracking photochemical reactions in the upper layers of the atmosphere. Given that our target planets orbit stars with different effective temperatures, they receive varying UV fluxes, which significantly influence photochemical processes. Accordingly, we have applied distinct UV stellar flux profiles specific to the stellar effective temperatures of each planet - (1) HAT-P-11 b: `\texttt{sflux-epseri}', (2) HD 63433 c: `\texttt{Gueymard-solar}', (3) K2-25 b: `\texttt{sflux-GJ1214}', and (4) TOI-270 d: `\texttt{sflux-GJ436}'. These flux profiles were taken from the VULCAN repository and represent stars of similar spectral classes with comparable effective temperatures. We adopted a vertically constant eddy diffusion coefficient, $K_{\mathrm{zz}}$ = $\mathrm{10^{10}}$ $\mathrm{cm^2/s}$, for all atmospheres, following the most common value applied by \cite{Tsai_2017} and \cite{2024ApJ...972..165D}. For all the planets, we utilized the default `\texttt{SNCHO full photo network}', which includes 1286 reactions. The host star radius and star-planet distance, as listed in Table \ref{tab:system}, were used to determine the stellar flux at the star's surface and convert it to the flux received by the planet. This conversion was performed on a uniform grid using the trapezoidal integral method inside the model. Additionally, the planet's radius and gravity are crucial for estimating the atmospheric scale height and initiating atmospheric dynamics.

\subsection{High-resolution opacity calculation for molecules}
\label{sec:opacity}

High-resolution molecular opacity is crucial for radiative transfer calculation of the planetary atmosphere and estimating the detectability of molecular bands. petitRADTRANS has its unique format for opacity files, spanned over a wide 130 pressure-temperature grid points (13 grid points for temperature: [81 K, 110 K, 148 K, 200 K, 270 K, 365 K, 493 K, 666 K, 900 K, 1215 K, 1641 K, 2217 K, and 2995 K] and 10 grid points for pressure: [$\mathrm{10^{-6}}$ to $\mathrm{10^{3}}$ bar in log-scale]). We follow the instructions from the petitRADTARNS manual \footnote{\href{https://petitradtrans.readthedocs.io/en/latest/content/adding\_opacities.html}{https://petitradtrans.readthedocs.io/en/latest/content/adding\_\\opacities.html}} to calculate the molecular opacity files at a resolution of $\mathrm{10^{6}}$. We use ExoCross\footnote{\href{https://exocross.readthedocs.io/en/latest/}{https://exocross.readthedocs.io/en/latest/}} \citep{yurchenko2018exocross} for the same. ExoCross is a Fortran-based tool that is capable of computing thermodynamic properties from molecular line lists. We use the line lists from ExoMol \citep{tennyson20202020} and HITRAN \citep{gordon2022hitran2020} as input files for ExoCross to generate high-resolution molecular opacity files for petitRADTARNS-specific pressure-temperature grid points.

The ExoMol line lists are distributed among three file structures: (1) a transition file (.trans), (2) a states file (.states), and (3) a partition function file (.pf). The ``transition" files contain the upper state ID, lower state ID, and the Einstein coefficients between those states, respectively. On the other hand, the ``states" file comprehensively details the rotational-vibrational (rovibronic) states, encompassing their energy in $\mathrm{cm^{-1}}$, collective degeneracy, energy uncertainty in $\mathrm{cm^{-1}}$, total angular momentum, lifetimes, and g-lambda factors. Additionally, it includes the associated quantum numbers, with each entry uniquely identified by a State ID (an integer). Each transition in the HITRAN database (.par file) includes essential details such as the line position, intensity, lower-state energy, line shape parameters, and assignment information. These details are crucial for calculating the absorption of a transition at specific temperatures and concentrations. HITRAN's line-by-line parameters are given at a standard reference temperature of 296 K. The intensities are adjusted to reflect the usual natural abundance found on Earth. During opacity calculations, the default pressure broadening (width of the Lorentz profile in $\mathrm{cm^{-1}}$) is applied to molecules lacking specific broadening parameters. For other molecules, $\mathrm{H_{2}}$ - He broadening is considered. This setting is tailored for an atmosphere with a primordial composition \textbf{(low mean molecular weight (MMW) atmosphere)}, where $\mathrm{H_{2}}$ and He constitute approximately 86\% and 14\% of the atmosphere, respectively. \textbf{This particular assumption for broadening coefficients may not fully capture the line-broadening effects in high MMW atmospheres that are not $\mathrm{H_{2}}$ - He dominated. While this assumption provides a reasonable approximation for a wide range of atmospheric conditions, it could introduce some uncertainties in the model calculations for such high MMW atmospheres.}

A summary of the opacity generation process is presented in Figure \ref{fig:flowchart}. The input file includes the names of the line list files, wavenumber limits (39 - 91000 $\mathrm{cm^{-1}}$), molecular mass, pressure-temperature grid, and broadening parameters. Once high-resolution opacity is calculated for the specified pressure-temperature grids, the opacity files are rebinned to the high-resolution wavelength grid points used by petitRADTRANS. Subsequently, a high-resolution opacity conversion routine is applied using the specified wavelength limits (0.3-28 $\mu$m) and molecular mass to convert the line-by-line opacity tables from ExoCross to petitRADTRANS format. \textbf{The conversion routine extracts opacity values within a user-specified wavelength range and normalizes the opacity data by dividing it with the molecular mass. This conversion restructures high-resolution opacity data from ExoCross into a wavelength-filtered and molecular-mass-normalized format compatible with petitRADTRANS.} The molecules for which high-resolution opacity calculations are performed, along with the respective databases used, are listed in Table \ref{tab:opacity}.

\begin{table}
  \centering
   \caption{Molecules and the respective database used to calculate the high-resolution line-by-line (R = $\mathrm{10^{6}}$) opacities in the petitRADTRANS grid. }
  \begin{tabular}{ccc} 
  \hline \hline
  \noalign{\smallskip}
     Molecules & Line lists & References\\
     \noalign{\smallskip}
     \hline
     \noalign{\smallskip}
     SO & ExoMol & \textbf{\cite{brady2024exomol}}\\ 
     CS & ExoMol & \textbf{\cite{paulose2015exomol}}\\
     SH & ExoMol & \textbf{\cite{gorman2019exomol}}\\
     SCO & HITRAN & \textbf{\cite{HITRAN2020}}\\ 
     SN & ExoMol & \textbf{\cite{yurchenko2018exomol}}\\
     $\mathrm{CS_2}$ & HITRAN & \textbf{\cite{CS2-gamma_air-1-1299}}\\ 
     \noalign{\smallskip}
     \hline
     \hline
  \end{tabular}

  \label{tab:opacity}
\end{table}


\begin{figure*}
    \centering  
        \begin{minipage}[b]{\columnwidth}
            \includegraphics[width=\columnwidth]{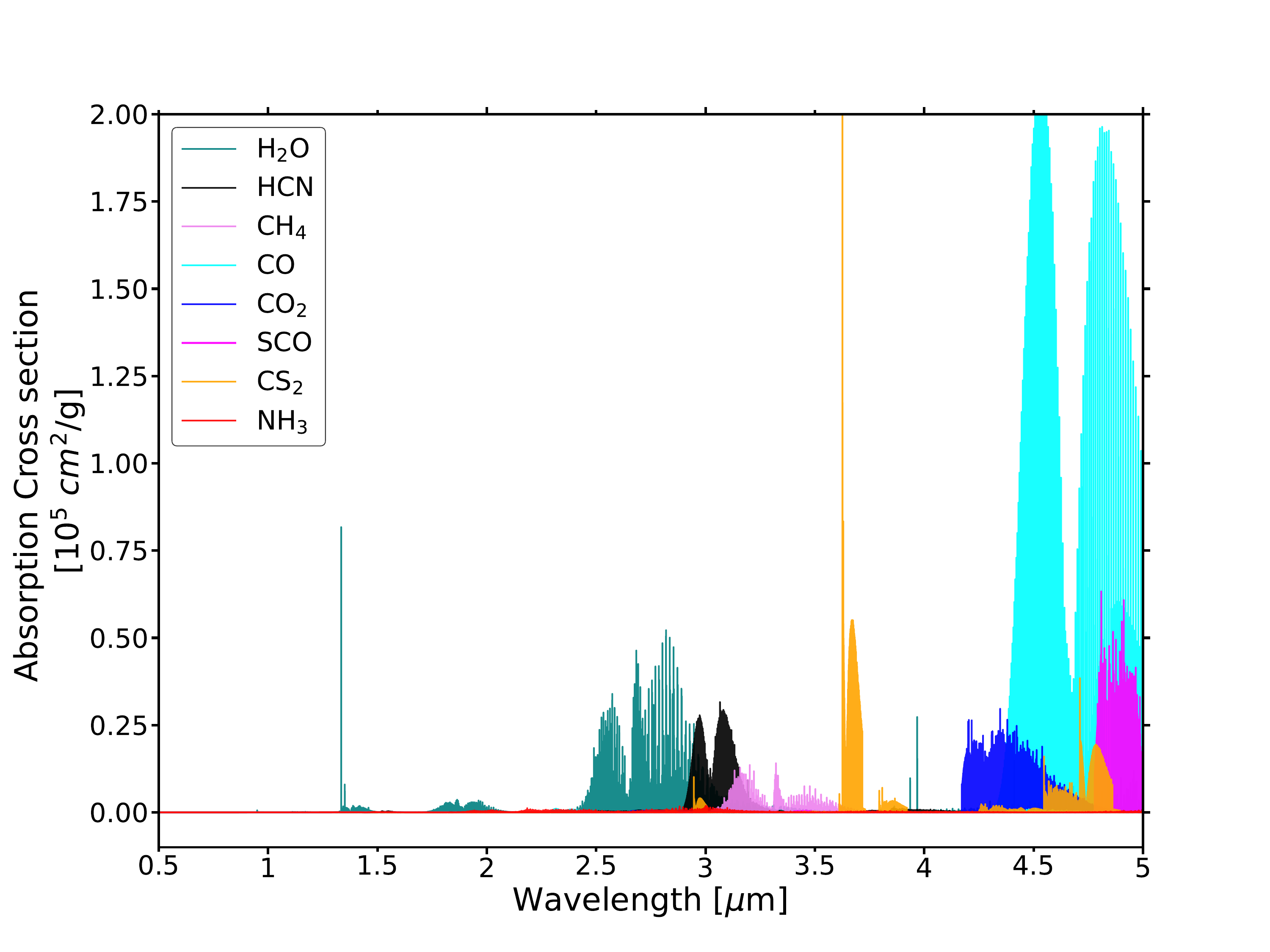}     
        \end{minipage}
        \begin{minipage}[b]{\columnwidth}
            \includegraphics[width=\columnwidth]{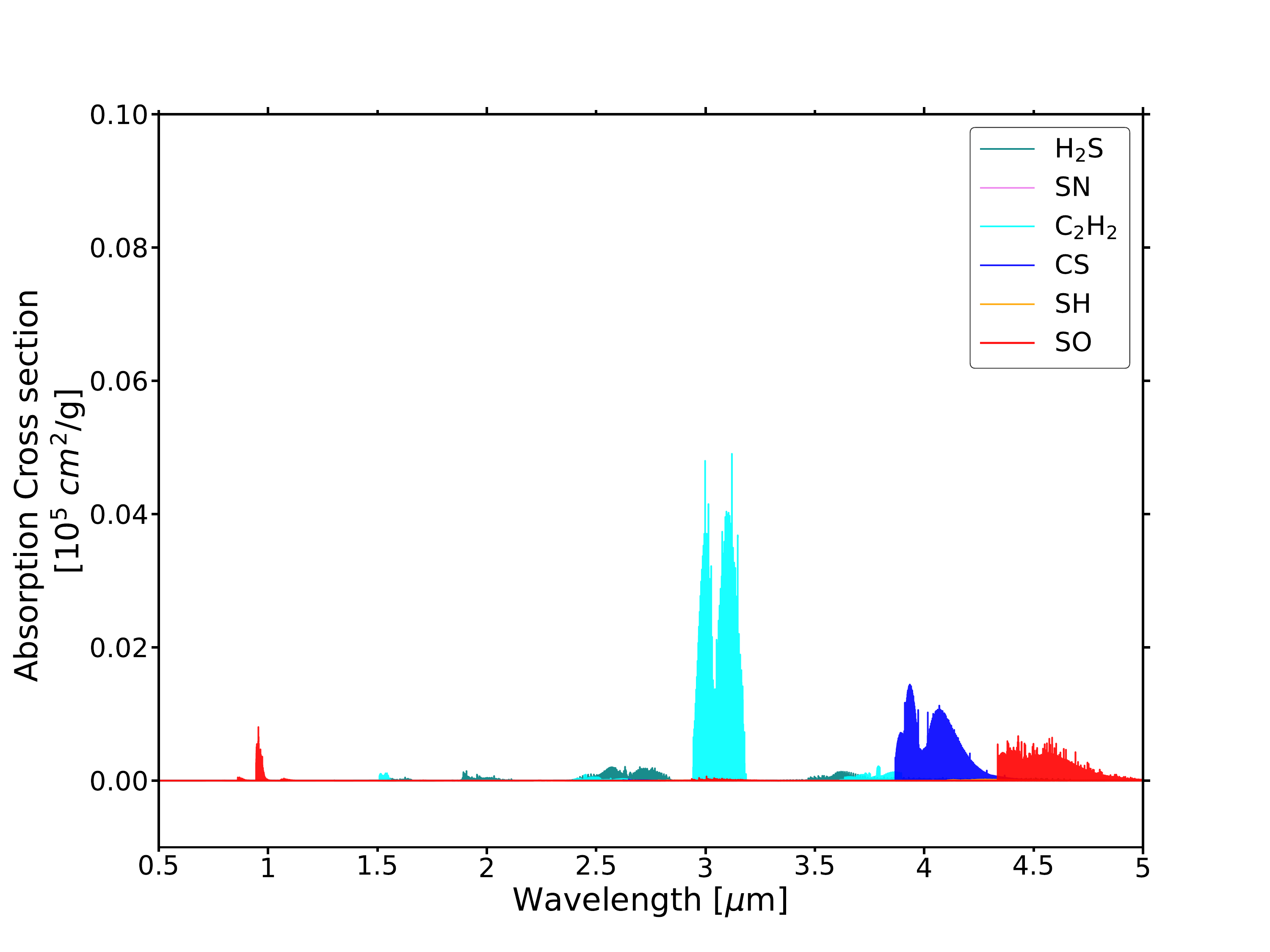}
        \end{minipage}
\caption{\textbf{High-resolution cross-section for key species of interest plotted as a function of wavelength at 1641 K and $\mathrm{10^{-2}}$ bar }}
\label{fig:sensitivity_and_opacity}
\end{figure*}


\begin{table*}
\centering
\begin{threeparttable}
\caption{Architecture and instrumental parameters for six current and future ground-based spectrographs}
\begin{tabular*}{\textwidth}{@{\extracolsep{\fill}}lcccccc@{}}

\hline
\hline
\noalign{\smallskip}
Telescope/ & Wavelength & Resolution & Diameter [m] & Read Noise & Dark Current  & $\mathrm{T_{opt}}$\\
Instrument & [$\mu$m]& & [m] & [$\mathrm{e^{-}}$] & [$\mathrm{e^{-}}$/pixel/s] & [K]\\
\noalign{\smallskip}
\hlineB{3.5}
\noalign{\smallskip}
TNG (GIANO-B) & 0.95-2.45 & 50000 & 3.58 & 5 & 0.05 & 50\\
\noalign{\smallskip}
CAHA (CARMENES) & 0.52-1.71 & 80000 & 3.5 & 13.52 & 0.002 & 140 \\
\noalign{\smallskip}
Gemini (IGRINS) & 1.45-2.45 & 45000 & 8.1 & 5 & 0.01 & 77 \\
\noalign{\smallskip}
Keck (HISPEC) & 0.98-2.46 & 100000 & 10 & 3 & 0.015 & 143 \\
\noalign{\smallskip}
TMT (MODHIS) & 0.95-2.5 & 100000 & 30 & 3 & 0.0002 (optical) & 143 \\
\noalign{\smallskip}
& & & & & 0.015 (NIR) &\\
\noalign{\smallskip}
E-ELT (ANDES) & 0.4-1.8 & 100000 & 39.3 & 3 & 0.00056 (optical) & 143 \\
& & & & & 0.0011 (NIR) &\\
\noalign{\smallskip}

\hline
\hline
\label{tab:instruments}
\end{tabular*}
\begin{tablenotes}[flushleft]
        \item[a] Note: The instrumental configurations are taken from the respective websites of the high-resolution spectrographs. GIANO-B: \href{https://www.tng.iac.es/instruments/giano-b/}{https://www.tng.iac.es/instruments/giano-b/}, CARMENES: \href{https://www.caha.es/telescope-3-5m/carmenes}{https://www.caha.es/telescope-3-5m/carmenes}, IGRINS: \href{https://www.gemini.edu/instrumentation/igrins}{https://www.gemini.edu/instrumentation/igrins}, HISPEC: \href{https://etlab.caltech.edu/instruments/hispec}{https://etlab.caltech.edu/instruments/hispec}, MODHIS: \href{https://www.tmt.org/page/modhis}{https://www.tmt.org/page/modhis}, and ANDES: \href{https://elt.eso.org/instrument/ANDES/}{https://elt.eso.org/instrument/ANDES/}.
    \end{tablenotes}
    \end{threeparttable}
\end{table*}

\subsection{Synthetic observations: coupling of line-by-line radiative transfer with simulated noise}
\label{sec:noise}

We simulate high-resolution transmission spectra for each planet by integrating time-evolved chemistry from VULCAN with the radiative transfer model petitRADTRANS\footnote{\href{https://petitradtrans.readthedocs.io/en/latest/}{https://petitradtrans.readthedocs.io/en/latest/}} and performing line-by-line radiative transfer at a reference pressure layer of $P_0$ = 0.01 bar. Two atmospheric conditions are considered in our case: (1) a cloud-free/clear atmosphere, and (2) a cloudy model that includes a gray cloud deck at 0.01 bar and a haze-like component, simulated by scaling the Rayleigh scattering of the gas by a factor of 10. The use of the cloud deck aims to explore the variability in molecular detection across different planetary atmospheres. \textbf{However, the detection or non-detection of molecules in transmission spectra depends on other factors such as disequilibrium chemistry, atmospheric abundances, and composition, etc. By comparing clear and cloudy atmospheres, this study isolates and highlights the specific role of clouds as one of the potential contributors to variability in molecular detection.} The molecular species used in the radiative transfer calculations for each planet are listed in Tables \ref{tab:detetcion_hatp11b}, \ref{tab:detetcion_hd63433c}, \ref{tab:detetcion_k225b}, and \ref{tab:detetcion_toi270d}. These opacities are sourced from the petitRADTRANS high-resolution opacity database or calculated as described in Section \ref{sec:opacity}.\textbf{ Figure \ref{fig:sensitivity_and_opacity} provides the high-resolution cross-sections of key species at 1641 K and $\mathrm{10^{-2}}$ bar.} Molecule selection is based on the VMR profiles, specifically those with an average VMR in the photosphere region (between 1 bar and 0.1 mbar) greater than $10^{-8}$. Molecules with lower VMRs are considered insignificant due to their negligible contributions to the overall planetary transmission spectra. Additionally, collision-induced absorption (CIA) opacities for $\mathrm{H_2}$-$\mathrm{H_2}$ and $\mathrm{H_2}$-He, as well as Rayleigh scattering by $\mathrm{H_2}$ and He, are included. The default resolution of the line-by-line opacity files is $10^6$. \textbf{To match the synthetic spectra with the spectrograph resolutions, we utilize the inbuilt opacity sampling method of petitRADTRANS, which rebins the opacity files from the default resolution to the instrument-specific resolutions listed in Table \ref{tab:instruments}.}

\textbf{We use the opacity sampler in petitRADTRANS to rebin opacities from their intrinsic resolution of R = $10^6$ to the desired instrument resolutions ($\mathrm{R_{i}}$). This is achieved by specifying the appropriate downsampling factor (R/$\mathrm{R_{i}}$), which ensures that the opacity grids were precomputed to the required wavelength binning for each spectrograph. Additionally, we account for the instrument's line spread function (LSF), which describes how finely the light of a given frequency is resolved. Assuming a Gaussian LSF with standard deviation $\mathrm{\sigma_{LSF} = R/(2\sqrt{ln2}})$, the opacity-sampled spectra are convolved with the LSF before being rebinned to the wavelength grid of each spectrograph. This convolution step makes sure that the model resolution accurately reflects the instrument's ability to resolve spectral features, avoiding aliasing or undersampling effects. Neglecting this convolution process or inadequate sampling of the LSF can result in underestimated line broadening and over-optimistic predictions of molecular detectability. To address this, the wavelength binning of the model is Nyquist-sampled for all instruments by maintaining $\mathrm{\lambda/\Delta\lambda}$ $>$ $\mathrm{2\lambda/\Delta\lambda_{LSF}}$ (where $\mathrm{\lambda}$ = the wavelength at which the opacities in petitRADTRANS are defined, $\mathrm{\Delta\lambda}$ = the wavelength difference between two neighboring wavelength points and $\mathrm{\Delta\lambda_{LSF}}$ = the FWHM of the spectrograph’s LSF), as outlined in the petitRADTRANS documentation. This approach preserves the fidelity of our simulations by properly accounting for the effects of resolution and detector properties.}

After completing the radiative transfer calculations, the next objective is to incorporate noise into the spectra to simulate synthetic observations. Ground-based observations are more challenging than space-based ones due to the contamination from Earth's telluric lines. When conducting ground-based observations, it is essential to consider the telluric transmittance and separate the planetary signal from the observed signal. This challenge can be addressed by applying Doppler shifts to the planetary spectrum and simulating both out-of-transit and in-transit observations, with the telluric lines included in the simulation. In addition to the Earth's telluric component, various other conventional noise sources (such as stellar noise, background radiation, airglow, moonlight, and thermal emission) must be considered during the observations. The model SPECTR\footnote{\href{https://github.com/curriem/spectr}{https://github.com/curriem/spectr}} is a ground-based, coronagraph-free noise simulator pipeline designed to account for these diverse noise sources. It simulates observations and assesses the detectability of molecular features using the cross-correlation spectroscopy technique. SPECTR simulates the wavelength- and resolution-dependent telluric lines using the Cerro Paranal Advanced Sky Model (SkyCalc)\footnote{\href{https://www.eso.org/observing/etc/doc/skycalc/helpskycalccli.html}{https://www.eso.org/observing/etc/doc/skycalc/helpskycalccli.html}} \citep{noll2012atmospheric}, developed by the European Southern Observatory (ESO). For our simulations, we assume the observatory's location at Paranal, with an atmospheric precipitable water vapor content of 3.5 mm. Detailed parameters specific to instrument resolution, wavelength coverage, read noise, and dark current are provided in Table \ref{tab:instruments}. Building on the gas giant observations conducted using CARMENES \citep{landman2021detection, sanchez2022searching, dash2024constraints} and the recent modeling work by \cite{2024ApJ...972..165D}, \textbf{we have assumed a fixed exposure duration of 500 seconds across all targets as part of the SPECTR framework, which simplifies the analysis by treating the exposure time as equal to the approximate transit duration.} Our analysis incorporates both singular transit observations and repeated transits, totaling three occurrences. This approach facilitates a comparative analysis of the detectability of molecular features across different spectrographs.

\textbf{The instrument efficiencies are maintained consistently across their respective wavelength ranges:  21\% for GIANO-B \citep{oliva2006giano}, 7.9\% for CARMENES \citep{seifert2016carmenes}, 6\% for IGRINS \citep{LE20152509}, and 10\% for HISPEC \citep{Mawet2019High}, MODHIS and ANDES \citep{currie2023there} respectively. Although throughput typically varies with wavelength, our assumption of uniform throughput serves to provide a conservative estimate of molecular detection limits.} Mirror temperatures are held constant at 273 K, following \cite{currie2023there} and \cite{2024ApJ...972..165D}, while detector temperatures are specified according to Table \ref{tab:instruments} in the respective instrument specifications. The synthetic observations require essential star-planet parameters, including the system's distance from Earth, systematic radial velocity, star's rotation period (we consider a period of 10 days for TOI-270 star following the default value from SPECTR repository), planet's orbital period, semi-major axis, and transit duration, detailed in Table \ref{tab:system}. We consider a constant barycentric radial velocity of 2 km/s for all planets and 90$\mathrm{^o}$ inclination for transmission. \textbf{A uniform detector pixel count of 100 per resolution element is maintained for all instruments as part of the SPECTR framework.} Our data processing pipeline incorporates an outlier mask and high-pass filters to mitigate low signal-to-noise ratio (S/N) data points and suppress low-frequency variations effectively.

Finally, the model employs cross-correlation spectroscopy to assess the detectability ($\mathrm{\sigma_{det}}$) of molecular bands from planet spectra. Following a method outlined in \cite{currie2023there} and \cite{2024ApJ...972..165D}, we cross-correlate the simulated spectra ($\mathrm{f(\lambda)}$) of planetary atmospheres with Doppler-shifted template spectra ($\mathrm{f(\lambda - \lambda')}$) across a range of Doppler velocities (-150 km $\mathrm{s^{-1}}$ to +150 km $\mathrm{s^{-1}}$). Key to our approach is computing the cross-correlation function (CCF), which peaks at zero relative velocity when a molecular feature is detectable. The CCF is derived from the covariance between observed and template spectra, normalized by their respective variances. Subsequently, a $\mathrm{\chi^2}$ analysis compares the CCF to a flat line, focusing on velocity shifts ($|\lambda|$) less than 5 km $\mathrm{s^{-1}}$, typical of planetary signals. This analysis yields a \textit{p}-value, which is converted into a $\mathrm{\sigma_{det}}$, indicating the deviation of the CCF from a Gaussian distribution. This method demonstrates robustness against minimal radial velocity perturbations and is applicable across various ground-based facilities operating at different resolutions and wavelength regimes, offering a sensitive tool for detecting molecular signatures in exoplanetary atmospheres using a high-resolution ground-based spectrograph. A detailed mathematical approach is outlined in \cite{brogi2019retrieving}.


\subsection{\textbf{Limitations of SPECTR in realistic observational scenarios}}
\label{sec:limitations}

\textbf{While SPECTR provides a simplified and idealized framework for studying the detectability of exoplanetary atmospheres using high-resolution spectroscopy, it has several caveats and limitations compared to the methodologies currently employed in the field. The assumptions made in SPECTR, such as uniform exposure times, standardized noise models, and fixed resolution elements, are designed to streamline the analysis and provide optimistic estimates of detectability. However, these assumptions do not fully reflect the complexities of real-world observations, including variability in stellar brightness, instrumental configurations, and advanced techniques such as phase-dependent corrections and likelihood-based detection frameworks. Below, we outline the key limitations of SPECTR and their implications for realistic observational scenarios.}

\begin{figure*}
    \centering  
        \begin{minipage}[b]{\columnwidth}
            \includegraphics[width=\columnwidth]{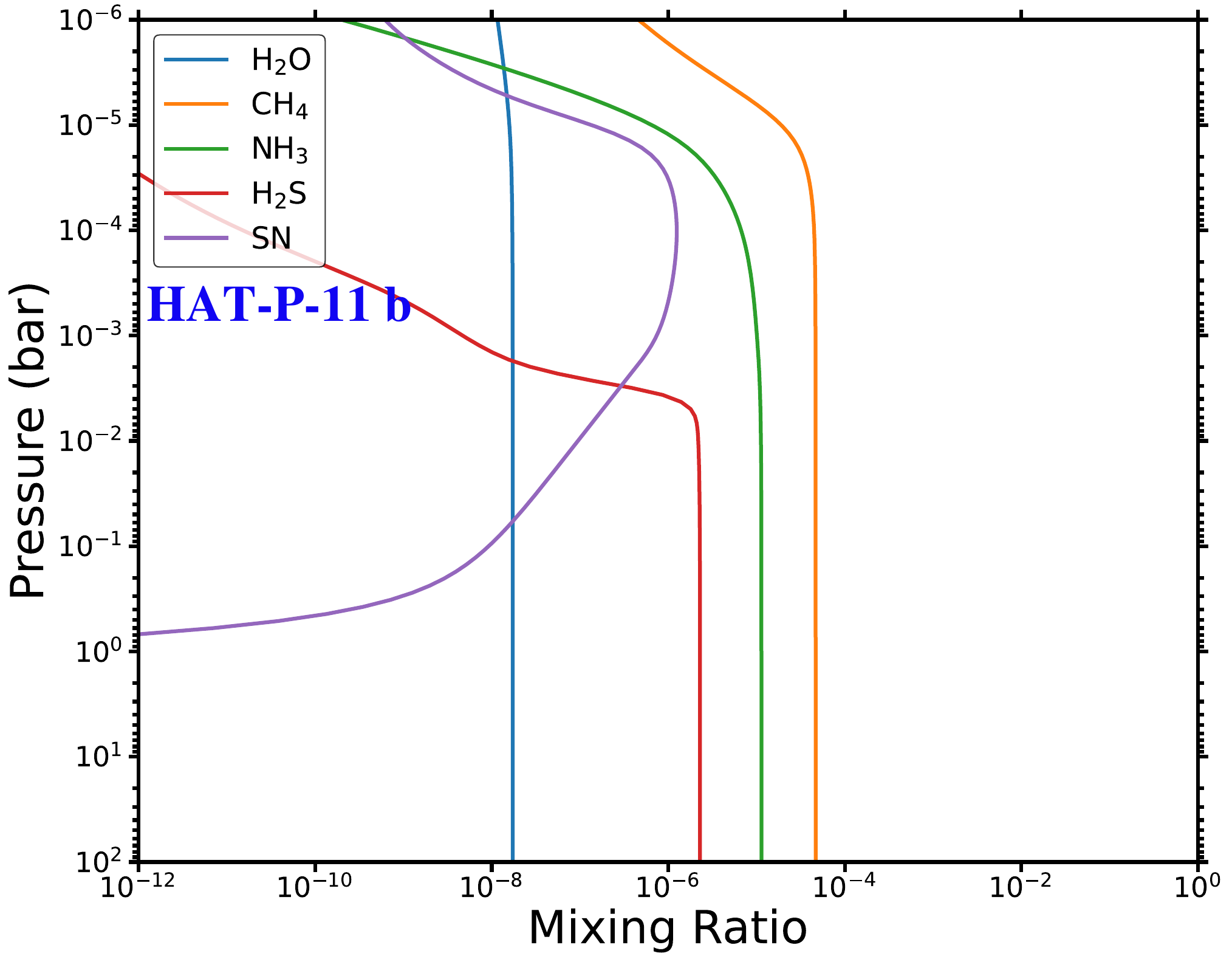}
        \end{minipage}
        \begin{minipage}[b]{\columnwidth}
            \includegraphics[width=\columnwidth]{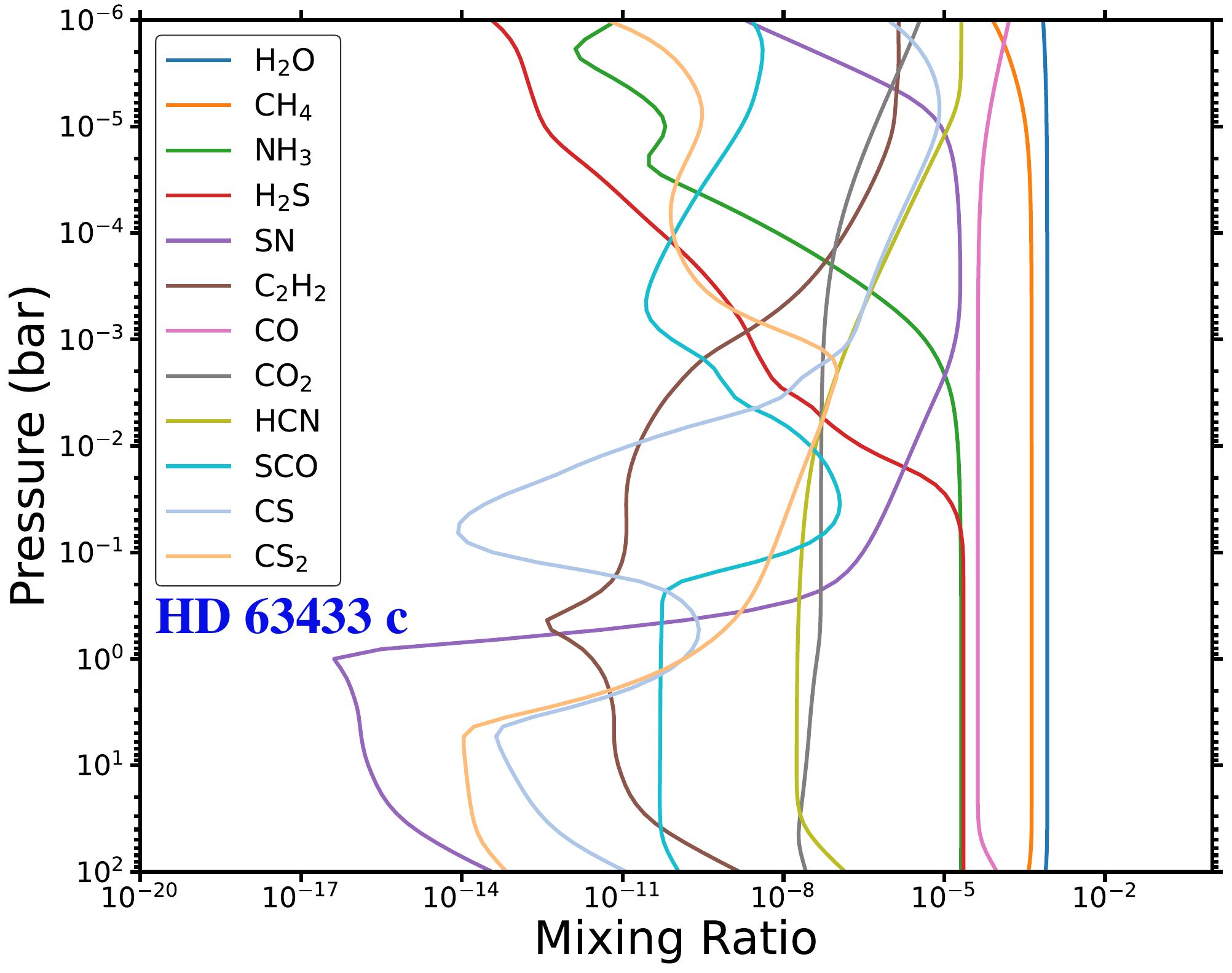}
        \end{minipage}
        \begin{minipage}[b]{\columnwidth}
            \includegraphics[width=\columnwidth]{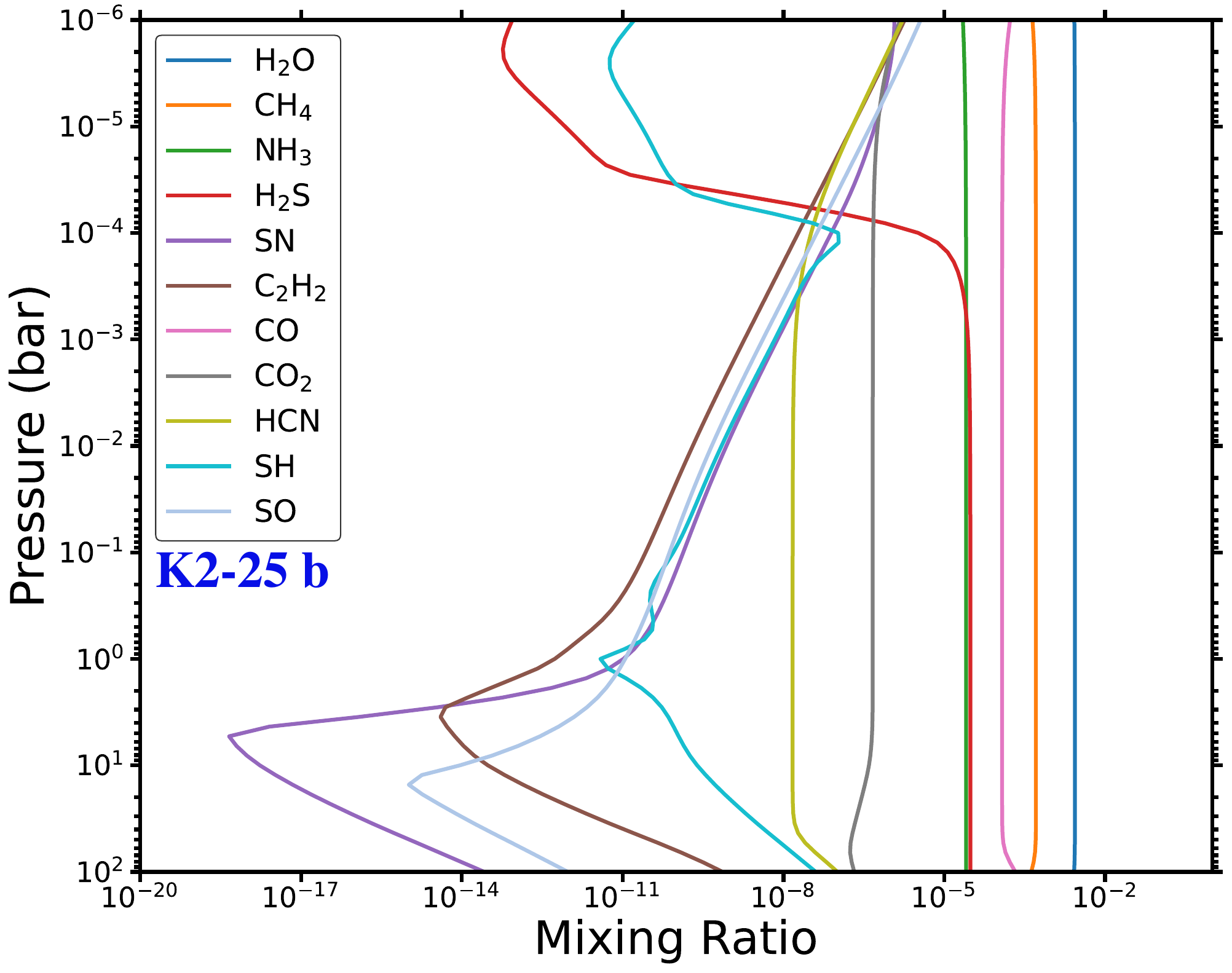}
        \end{minipage}
         \begin{minipage}[b]{\columnwidth}
            \includegraphics[width=\columnwidth]{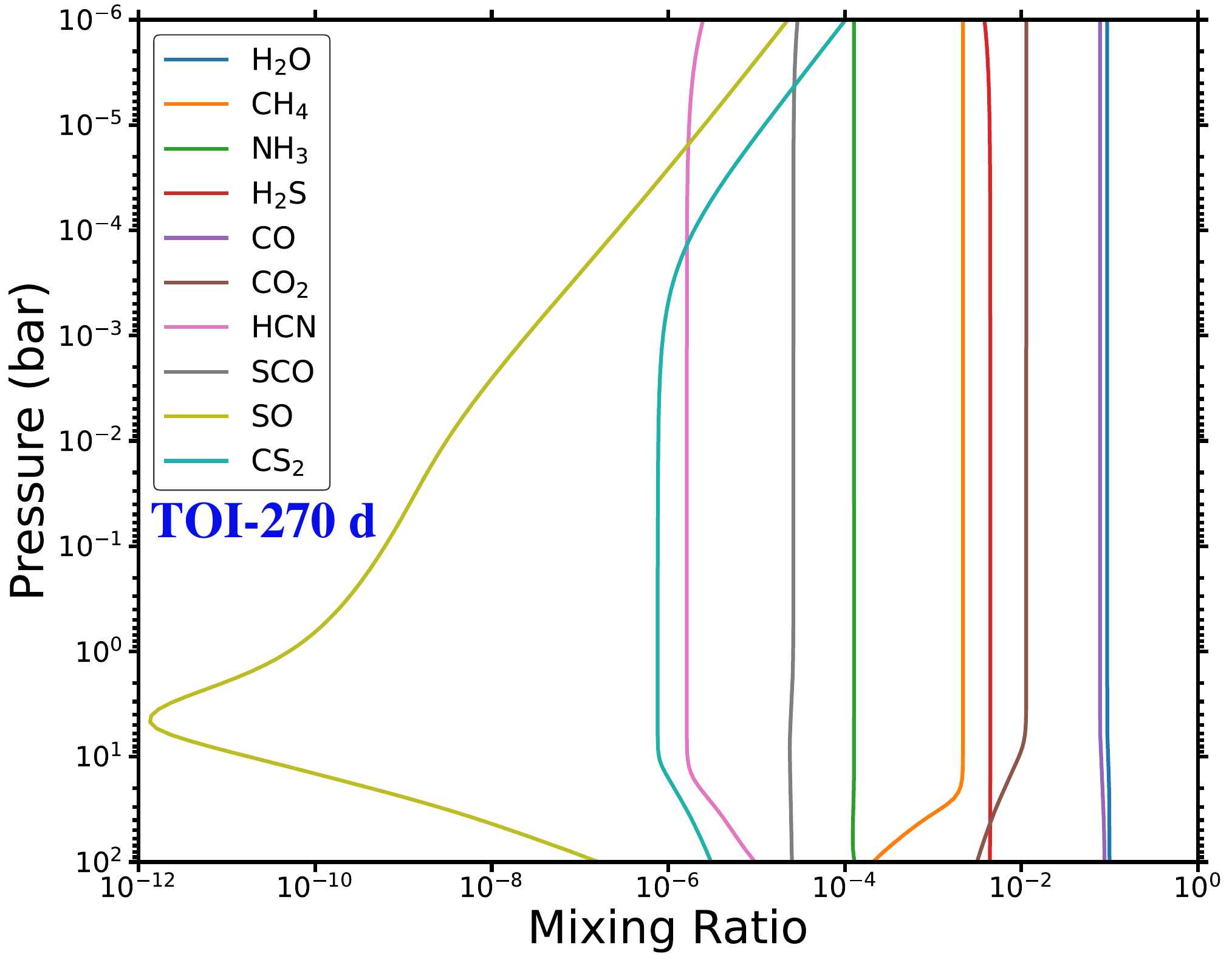}
        \end{minipage}
\caption{Volume mixing ratio profiles for selected molecules in the atmospheres of four planets. Molecules were chosen based on having an average photospheric mixing ratio greater than $\mathrm{10^{-8}}$. The pressure layers defining the photosphere boundary are 1 bar and 0.1 mbar in our case.}
\label{fig:VMR}
\end{figure*}

\begin{figure*}
    \centering  
        \begin{minipage}[b]{\columnwidth}
            \includegraphics[width=\columnwidth]{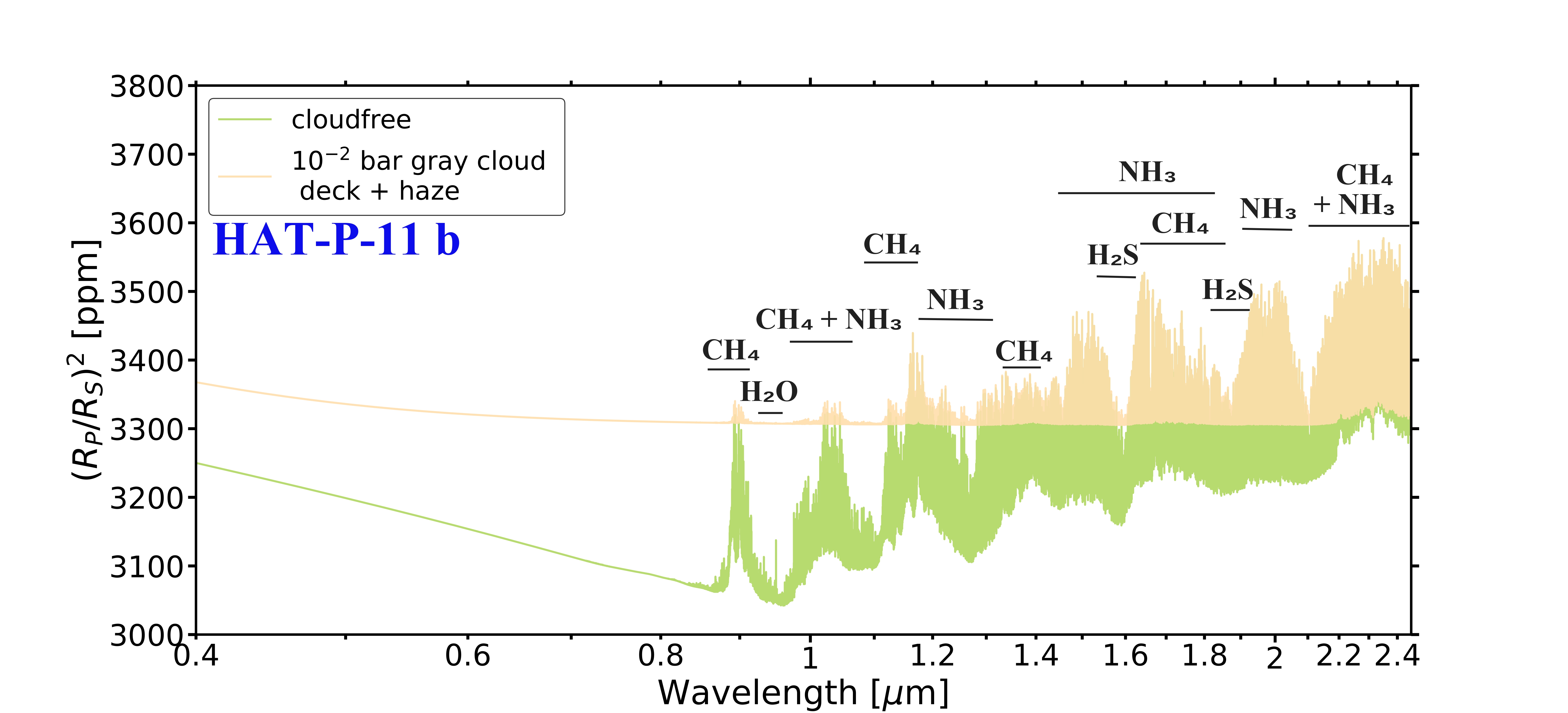}
        \end{minipage}
         \begin{minipage}[b]{\columnwidth}
            \includegraphics[width=\columnwidth]{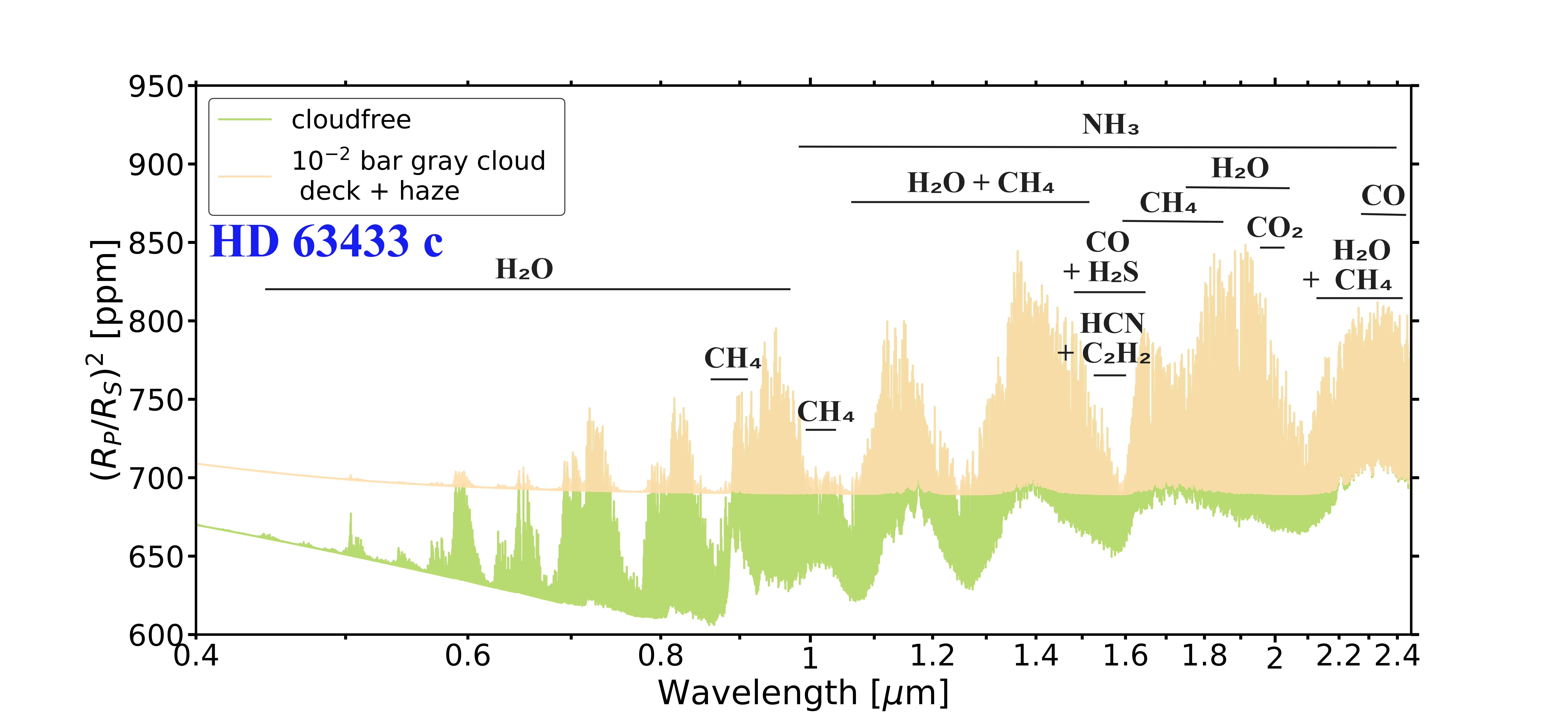}
        \end{minipage}
         \begin{minipage}[b]{\columnwidth}
            \includegraphics[width=\columnwidth]{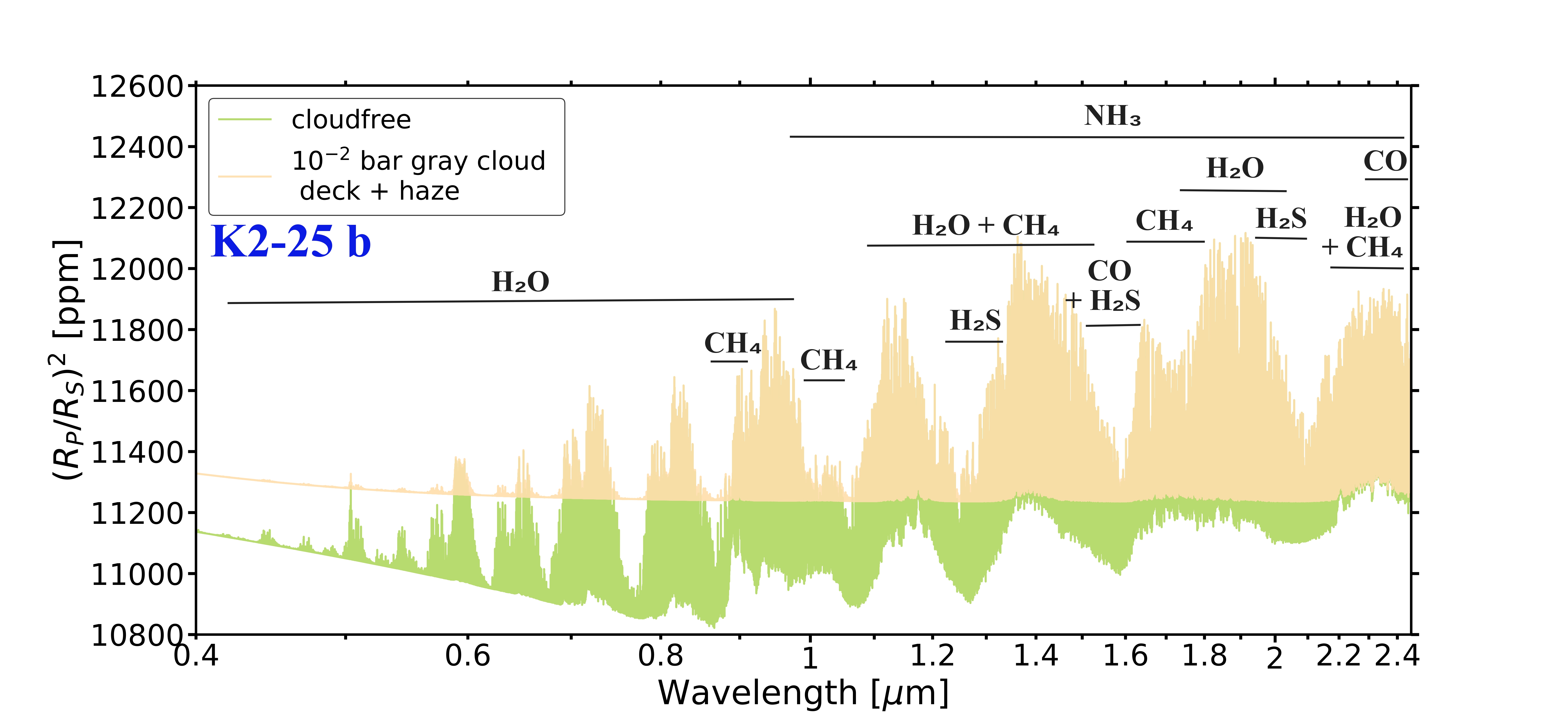}
        \end{minipage}
        \begin{minipage}[b]{\columnwidth}
            \includegraphics[width=\columnwidth]{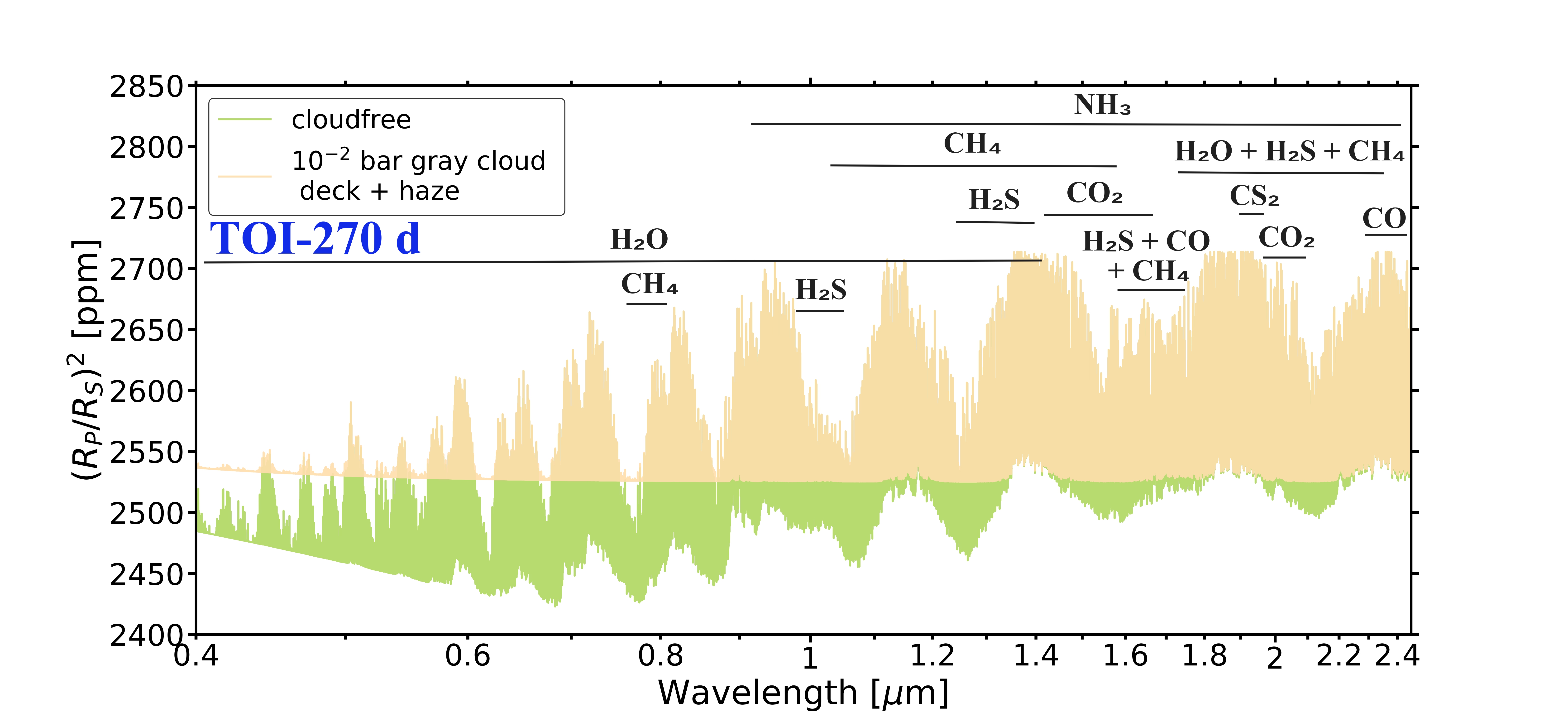}
        \end{minipage}
\caption{High-resolution transmission spectra for four planets, with a resolution akin to GINAO-B (R = 50,000) and full wavelength coverage for all spectrographs (0.4-2.45 $\mu$m). Each plot illustrates the spectra under two atmospheric conditions: clear (depicted in green) and cloudy (depicted in gold). The cloud model includes a gray cloud deck at 0.01 bar with a haze layer enhancing Rayleigh scattering scaled by a factor of 10. The band-wise contributions from individual molecules to the overall planetary spectra are also depicted.}
\label{fig:Transmission_spectra}
\end{figure*}

\textbf{We have considered the same exposure time for all the targets in our analysis. While this approach is designed to provide an optimistic prediction of detectability, it does not fully account for target-specific observational complexities. For example, stellar hosts with significantly varying brightness, such as TOI-270 ($\mathrm{K_{mag}}$ = 8.2) \citep{gunther2019super} and HD 63433 ($\mathrm{K_{mag}}$ = 5.3) \citep{brown2018gaia}, could experience either pixel saturation for brighter stars or low S/N for fainter stars. Similarly, long exposure times for fast-moving planets could lead to Doppler smearing of planetary lines, reducing signal coherence during integration \citep{rasmussen2023nondetection}. Although such effects can be mitigated by adjusting exposure durations or correcting templates for smearing, they are beyond the scope of this study. Our analysis also assumes that all exposures are taken under consistent airmass and precipitable water vapor conditions and that the telluric line removal techniques, such as Principal Component Analysis (PCA) based methods \citep{boldt2024optimising}, performs ideally under these conditions resulting in perfect detrending with no imprint left on the exoplanet signal. While these assumptions simplify the analysis, they do not fully reflect the variability or complexities in real observations. Instead, we focus on an analysis where the effects of smearing, saturation, airmass, and water vapor variation can be neglected, providing an idealized framework to evaluate detectability. Future work could expand upon this analysis by incorporating target-specific observational strategies, telescope and instrument characteristics, and corrections for these effects to better reflect real-world scenarios.}

\textbf{Another key limitation of the SPECTR framework lies in assuming a constant detector pixel count of 100 per resolution element for all instruments. This assumption provides a standardized comparison across all simulations, allowing us to assess detectability under controlled conditions. The wavelength-dependent S/N calculation count per resolution element includes contributions from photon noise, readout noise, and dark current, with the total noise modeled following the methodology outlined in the SPECTR model. Specifically, photon noise is simulated by summing up the signal, background, and dark current for each resolution element, with the total shot noise calculated as the square root of this sum. The total noise is obtained by adding shot noise and readout noise in quadrature. Random values are then drawn from Poisson distributions defined by the total noise for each resolution element, ensuring that noise sampling varies across pixels. It is important to note that this approach does not assume uniform brightness for all stellar hosts. Instead, noise is modeled individually for each target, accounting for differences in stellar brightness, telescope throughput, and background contributions. While this idealized framework simplifies the noise treatment and standardizes resolution elements, it does not fully capture the complexities of real observational setups, where effective pixel counts can vary depending on instrument resolution. A higher pixel count per resolution element would lead to a smaller average S/N per resolution element assuming the wavelength range remains constant. Future work (as \cite{currie2023there} also mention in their SPECTR paper) could help refine these assumptions by incorporating realistic detector and instrument-specific parameters to better reflect the diversity of real-world observations.}

\textbf{Additional constraint in the SPECTR framework arises from its simplified assumptions in computing the 1D CCFs and determining detection significance, which differ from more robust methodologies commonly used in high-resolution spectroscopy studies. SPECTR adopts a simplified approach to compute the 1D CCFs, assuming a fixed radial velocity for the planet throughout the observation duration i.e. no time variability. While this approach is appropriate for cases where the planet’s orbital motion during the observation window is negligible or where the analysis focuses on a fixed radial velocity, we acknowledge that realistically the planetary motion along its Keplerian orbit introduces time-dependent velocity shifts. Hence, the conventional method for transmission spectroscopy involves computing a 1D CCF for each exposure by Doppler-shifting a model template spectrum by the planet’s radial velocity at the corresponding orbital phase and then cross-correlating with the (normalized) transit depth at that exposure. The peak of CCFs across exposures then traces the exoplanet’s orbit. Since an orbit can be constrained through a $\mathrm{K_p}$-$\mathrm{V_{sys}}$ map, where $\mathrm{K_p}$ is the Keplerian velocity semi-amplitude and $\mathrm{V_{sys}}$ is the systematic velocity \citep{brogi2012signature}, these two parameters together allow us to unambiguously associate the signal with the planetary origin by localizing a parameter space where the sum of the CCFs across exposures along an orbit is the highest.  Since SPECTR does not have the exoplanet moving along an orbit and instead has the signal injected at a fixed radial velocity at all exposures, a 1D CCF analysis as done in this study remains appropriate, and provides a best case estimate of the number of transits required to have enough S/N for an unambiguous detection of the injected exoplanet spectrum. Future work will aim to incorporate time variability to improve the robustness of results.} 

\textbf{The $\mathrm{\chi^2}$ function approach was chosen due to its simplicity and applicability in assessing how well a signal stands out from noise in regions outside the expected signal range. Specifically, it compares the variance from a baseline (assumed to be a non-detection) with the variance within the signal range, ensuring that $\mathrm{\sigma_{det}}$ $\sim$ 1 represents a null detection. While this approach has its limitations, it is consistent with the assumptions of the SPECTR model, which operates under an idealized scenario of perfect detrending. In this framework, in-transit spectra are divided by out-of-transit spectra (assuming that they are taken at the same distribution of airmasses) to isolate only the injected planetary signal and noise, avoiding potential artifacts introduced by detrending methods such as PCA or Singular Value Decomposition (SVD) \citep{brogi2019retrieving,dash2024constraints}. The other methods commonly used in high-resolution spectroscopy, such as the \textit{T}-test \citep{birkby2017discovery} or likelihood-based approaches \citep{brogi2019retrieving,gibson2020detection}, offer more robust frameworks for determining detection significance, particularly in cases where detrending processes may impact line shapes and depths. These methods often convert the CCF into a log-likelihood space and use Wilk's theorems to evaluate detection significance. While the $\mathrm{\chi^2}$ analysis does not capture line shapes and depths with the same fidelity, it remains suitable for our 1D analysis because the SPECTR model assumes no modification of the line profile during detrending. This assumption ensures that the observed line shapes remain consistent with the model template, making the $\mathrm{\chi^2}$ analysis adequate in our case.  Similar to \cite{currie2023there}, we adopt the less stringent (and more optimistic) $\mathrm{\sigma_{det}}$ $>$ 3 threshold to claim a detection, compared to the more stringent $\mathrm{\sigma_{det}}$ $>$ 4 used in other works in literature.}

\begin{sidewaystable*}
    
\vspace{9cm}
\begin{minipage}{1.05\textwidth}
\centering
\caption{Summary of $\mathrm{\sigma_{det}}$ achieved over one and three transits for HAT-P-11 b using various ground-based spectrographs. Values outside brackets indicate detection significance in a clear atmosphere, while values inside brackets denote detection significance in a cloudy atmosphere. Dashed values signify the instruments do not cover those specific bands, and `ND' represents a non-detection.}
\resizebox{\textwidth}{!}{
\begin{tabular}{lcc|cccccc|cccccc}
\\
\hline
\hline
\noalign{\smallskip}
&  & &   & & \multicolumn{2}{c}{Transit = 1} &  &   & & &  \multicolumn{2}{c}{Transit = 3} & \\
\noalign{\smallskip}
\hlineB{3.5}
\noalign{\smallskip}
Species & Bands & Wavelength  &  TNG & CAHA & Gemini & Keck & TMT & E-ELT &  TNG & CAHA & Gemini & Keck & TMT & E-ELT \\
 &  [$\mu$m] & range [$\mu$m]&  & & & & & & & & &  \\
\noalign{\smallskip}
\hlineB{3.5}
\noalign{\smallskip}
$\mathrm{H_2O}$ & 0.90 & 0.89 - 0.99 & ND & ND & -- & ND  & ND & 1.7 [ND] & ND & ND & -- & ND & 1.4 [ND] & 2.6 [ND] \\
& 0.94 & 0.93 - 0.95 & -- & ND & -- & --  &  -- & ND & -- & ND & -- & --  & -- & 2.5 [ND]\\
\noalign{\smallskip}
\hline
\noalign{\smallskip}
$\mathrm{CH_4}$ & 0.89 & 0.875 - 0.91 & -- & 3.9 [ND] & -- & --  &  -- & 15.2 [1.8] & -- & 4.2 [ND] & -- & --  & -- & 17.4 [3.6] \\
& 1.00 & 0.98 - 1.05 & ND & ND & -- & 4.0 [ND]  &  8.3 [ND] & 8.8 [1.4] & ND & ND & -- & 5.5 [ND]  &  10.9 [1.8] & 11.3 [3.3] \\
& 1.10 & 1.1 - 1.2 & 2.5 [ND] & 1.6 [ND] & -- & 2.7 [ND]  &  3.0 [1.5] & 9.8 [4.3] & 3.8 [1.2] & 1.9 [ND] & -- & 4.0 [1.3] &  5.0 [1.8] & 12.7 [8.4] \\
& 1.60 & 1.6 - 1.8 & 6.7 [3.6] & 6.0 [2.3] & 10.9 [10.3] & 9.7 [9.4]  &  15.6 [14.2] & 20.9 [17.8] & 10.9 [5.4] & 9.8 [5.0] & 13.4 [12.2] & 14.6 [14.2] &  16.2 [15.7] & 21.4 [19.5] \\
& 2.40 & 2.2 - 2.5 & ND & -- & ND & 9.4 [9.0]  &  19.5 [19.2] & -- & ND & -- & ND & 13.6 [10.7] &  21.8 [21.1] & -- \\
\noalign{\smallskip}
\hline
\noalign{\smallskip}
$\mathrm{NH_3}$ & 0.90 & 0.875 - 0.925 & -- & ND & -- & --  &  -- & 3.5 [1.4] & -- & ND & -- & --  &  -- & 6.7 [1.6] \\
& 1.00 & 1.0 - 1.1 & 5.8 [2.4] & 4.4 [ND] &  -- & 11.7 [2.5] &  16.7 [2.9] & 17.5 [3.1] & 9.8 [4.3] & 7.1 [2.0] & -- & 15.8 [4.5] &  18.0 [5.3] & 19.3 [5.9] \\
& 1.50 & 1.45 - 1.65 & 4.1 [3.4] & ND & 7.2 [4.4] & 6.8 [3.8]  &  9.6 [5.5] & 10.6 [10.0] & 6.1 [5.7] & ND & 8.7 [6.7] & 7.5 [6.5]  &  9.9 [9.4] & 16.0 [13.4] \\
& 2.00 & 2.0 - 2.1  & ND & -- & 2.1 [1.9] & 2.3 [2.0] &  5.3 [2.4] & -- & 2.4 [1.8] & -- & 3.2 [2.2] & 2.8 [2.4] &  9.5 [5.8] & -- \\
& 2.30 & 2.2 - 2.4 & 2.2 [2.0] & -- & 7.2 [6.9] & 9.2 [6.2]  &  13.7 [12.4] & -- & 3.7 [3.5] & -- & 10.2 [9.5] & 10.9 [9.4] &  16.7 [15.6] & -- \\
\noalign{\smallskip}
\hline
\noalign{\smallskip}
$\mathrm{H_2S}$ & 1.60 & 1.55 - 1.66 & ND  & ND & 2.6 [ND] & 2.7 [ND]  &  3.1 [ND] & 4.4 [1.7] & ND & ND & 3.8 [ND] & 4.2 [ND]  &  6.0 [ND] & 6.7 [2.6] \\
\noalign{\smallskip}
\hline
SN & all bands &  & ND & ND & ND & ND  &  ND & ND & ND & ND & ND & ND &  ND & ND\\
\noalign{\smallskip}
\hline
\hline
\noalign{\smallskip}
\label{tab:detetcion_hatp11b}
\end{tabular}}
\end{minipage}
\end{sidewaystable*}




\begin{sidewaystable*}

\vspace{-9cm}
\rotatebox{180}{
           \begin{minipage}{1.05\textwidth}
\centering
\caption{Similar to Table \ref{tab:detetcion_hatp11b} but for HD 63433 c.}
\resizebox{\textwidth}{!}{
\begin{tabular}{lcc|cccccc|cccccc}
\\
\hline
\hline
\noalign{\smallskip}
&  & &   & & \multicolumn{2}{c}{Transit = 1} &  &   & & &  \multicolumn{2}{c}{Transit = 3} & \\
\noalign{\smallskip}
\hlineB{3.5}
\noalign{\smallskip}
Species & Bands & Wavelength  &  TNG & CAHA & Gemini & Keck & TMT & E-ELT &  TNG & CAHA & Gemini & Keck & TMT & E-ELT \\
 &  [$\mu$m] & range [$\mu$m]&  & & & & & & & & &  \\
\noalign{\smallskip}
\hlineB{3.5}
\noalign{\smallskip}
$\mathrm{H_2O}$ & 0.90 & 0.89 - 0.99 & 12.2 [7.7] & 11.8 [5.4] & -- & 16.4 [12.4]  &  17.6 [16.4] & 18.3 [17.6] & 15.3 [9.6] & 13.3 [8.4] & -- & 17.1 [16.0]  &  17.8 [16.9] & 20.3 [20.0] \\
& 0.94 & 0.93 - 0.95 & -- & 7.5 [2.8] & -- & --  &  -- & 15.2 [14.5] & -- & 10.3 [5.0] & -- & --  &  -- & 16.0 [15.6] \\
& 1.00 & 0.95 - 1.05 & 11.6 [8.0] & 10.3 [7.1] & -- & 16.5 [12.7] &  17.5 [17.3] & 20.4 [19.6] & 13.0 [10.7] & 12.6 [7.5] & -- & 16.6 [15.7] &  17.6 [17.5] & 20.9 [20.2] \\
& 1.10 & 1.1 - 1.2 & 7.3 [6.2] & 5.8 [1.7] & -- & 7.8 [7.3] &  15.3 [8.7] & 16.4 [14.9] & 10.3 [7.9] & 8.0 [2.3] & -- & 11.2 [8.5] &  15.5 [9.8] & 17.8 [15.6] \\
& 1.30 & 1.3 - 1.55 & ND & ND & 9.0 [5.4] & ND  &  ND & ND & ND & ND & 12.9 [6.5] & ND  &  ND & ND \\
\noalign{\smallskip}
\hline
\noalign{\smallskip}
$\mathrm{CH_4}$ & 0.89 & 0.875 - 0.91 & -- & 4.8 [1.4] & -- & --  &  -- & 16.3 [12.8] & -- & 7.5 [2.4] & -- & --  &  -- & 18.7 [16.2] \\
& 1.00 & 0.98 - 1.05 & 4.3 [1.4] & 4.0 [1.1] & -- & 11.1 [2.1] &  16.7 [6.0] & 17.8 [6.9] & 6.8 [1.8] & 4.5 [1.7] & -- & 15.1 [5.9] &  19.0 [11.2] & 19.7 [12.2] \\
& 1.10 & 1.1 - 1.2 & 6.4 [2.5] & 3.4 [2.2] & -- & 6.8 [3.1] &  8.8 [6.0] & 13.5 [9.9] & 8.3 [7.3] & 4.5 [2.9] & -- & 8.5 [7.5] &  11.0 [8.1] & 16.6 [12.5] \\
& 1.30 & 1.3 - 1.55 & ND & ND & ND &  ND &  ND & ND & ND & ND & 1.5 [ND] & ND  &  ND & ND \\
& 1.60 & 1.6 - 1.8 & 10.4 [7.8] & 9.8 [6.1] & 13.3 [12.7] & 15.9 [14.3] &  16.6 [16.3] & 19.1 [17.3] & 12.1 [9.2] & 11.5 [8.2] & 14.7 [13.9] & 16.8 [14.9] &  17.1 [16.8] & 21.7 [18.9] \\
& 2.40 & 2.2 - 2.5 & 2.0 [ND] & -- & 10.8 [5.1] & 18.9 [18.6] &  22.6 [20.5] & -- & 5.3 [1.6] & -- & 13.3 [6.4] & 21.0 [19.5] &  23.0 [22.1] & -- \\
\noalign{\smallskip}
\hline
\noalign{\smallskip}
$\mathrm{C_2H_2}$ & 1.55 & 1.50 - 1.59 & ND & ND & ND & ND  &  ND & 4.3 [2.9] & ND & ND & 2.0 [ND] & ND  &  3.1 [ND] & 6.2 [3.2] \\
\noalign{\smallskip}
\hline
\noalign{\smallskip}
CO & 1.55 & 1.555 - 1.60 &  2.8 [ND] & ND & 3.9 [ND] & 4.2 [1.9] & 6.7 [3.1] & 7.2 [5.1] & 4.0 [2.9] & 1.9 [1.2] & 5.0 [3.1] & 6.6 [3.8] &  9.2 [5.7] & 9.5 [6.3] \\
& 2.30 & 2.3 - 2.5 & ND & -- & ND & 2.4 [ND] &  3.0 [2.1] & -- & ND & -- & ND & 2.9 [2.6] &  3.1 [2.8] & -- \\
\noalign{\smallskip}
\hline
\noalign{\smallskip}
$\mathrm{CO_2}$ & 2.00 & 1.94 - 2.08 & ND & -- & ND & ND  &   ND & -- & ND & -- & 2.1 [ND] & 1.9 [ND] &  3.7 [ND] & -- \\
\noalign{\smallskip}
\hline
\noalign{\smallskip}
HCN & 1.10 & 1.0 - 1.2 & ND & ND &  -- & ND  &  ND & ND & 1.4 [ND] & ND & -- & ND  &  1.5 [ND] & 2.1 [ND] \\
& 1.50 & 1.47 - 1.56 & ND & ND & 2.8 [2.1] & 3.4 [2.0] &  5.9 [5.2] & 7.0 [6.2] & 3.7 [2.6] & 1.7 [1.3] & 5.4 [4.7] & 5.2 [4.3] &  6.4 [6.9] & 7.7 [7.0] \\
\noalign{\smallskip}
\hline
\noalign{\smallskip}
$\mathrm{NH_3}$ & 0.90 & 0.875 - 0.925 & -- & ND & -- & --  &  -- & 3.8 [ND] & -- & 1.6 [ND] & -- &  --  & -- & 4.1 [ND] \\
& 1.00 & 1.0 - 1.1 & 8.0 [1.4] & 6.2 [ND] & -- & 13.9 [1.8] &  17.5 [4.6] & 17.8 [6.5] & 12.8 [1.8] & 8.4 [ND] & -- & 16.4 [3.3] &  18.1 [8.7] & 18.9 [10.4] \\
& 1.50 & 1.45 - 1.65 & 4.5 [3.0] & 1.5 [ND] & 5.9 [3.8]  & 5.3 [3.9] &  6.5 [4.5] & 11.9 [8.2] & 7.5 [4.8] & 3.1 [2.0] & 9.5 [6.7] & 9.1 [6.9] &  11.0 [7.3] & 12.9 [11.8] \\
& 2.00 & 2.0 - 2.1  & 4.7 [1.6] & -- & 5.1 [3.0] & 5.2 [3.1] &  7.2 [3.8] & -- & 5.4 [3.2]& -- & 7.2 [5.7] & 8.0 [6.0] &  11.4 [6.3] & -- \\
& 2.30 & 2.2 - 2.4 & ND & -- & 2.1 [1.9] & 2.8 [2.1] &  5.5 [3.8] & -- & 2.1 [1.6] & -- & 3.6 [3.2] & 3.5 [3.3] &  6.2 [4.9] & -- \\
\noalign{\smallskip}
\hline
\noalign{\smallskip}
SCO & all bands &  & ND & ND & ND & ND  &  ND & ND & ND & ND & ND & ND  &  ND & ND\\\noalign{\smallskip}
\hline
\noalign{\smallskip}
SN & all bands &  & ND & ND & ND & ND  &  ND & ND & ND & ND & ND & ND  &  ND & ND\\
\noalign{\smallskip}
\hline
\noalign{\smallskip}
CS & 2.15 & 2.0 - 2.3 & ND & -- & ND & ND  &  ND & -- & ND & -- & ND &  ND &  2.1 [2.0] & -- \\
\noalign{\smallskip}
\hline
\noalign{\smallskip}
$\mathrm{CS_2}$ & 1.88 & 1.875 - 1.89 & ND & -- & ND & ND  &  ND & -- & ND & -- & ND & ND  &  2.3 [ND] & -- \\
& 2.23 & 2.14 - 2.32 & ND & -- & ND & ND  &  ND & -- & ND & -- & 1.2 [ND] & 1.2 [ND] &  1.4 [ND] & -- \\
& 2.40 & 2.385 - 2.42 & ND & -- & ND & ND  &  ND & -- & ND & -- & ND & 1.4 [ND] &  1.6 [ND] & -- \\
\noalign{\smallskip}
\hline
\noalign{\smallskip}
$\mathrm{H_2S}$ & 1.60 & 1.55 - 1.66 & 1.7 [ND] & ND & 2.2 [ND] & 2.3 [ND] &  3.0 [ND] & 3.7 [1.8] & 2.6 [ND] & 2.0 [ND] & 4.0 [ND] & 3.9 [ND] &  4.2 [ND] & 5.9 [2.6] \\
\noalign{\smallskip}
\hline
\hline
\label{tab:detetcion_hd63433c}
\end{tabular}}
\end{minipage}}
\end{sidewaystable*}


\begin{sidewaystable*}
    
\vspace{9cm}
           \begin{minipage}{1.05\textwidth}
\centering
\caption{Similar to Tables \ref{tab:detetcion_hatp11b} and \ref{tab:detetcion_hd63433c} but for K2-25 b.}
\resizebox{\textwidth}{!}{
\begin{tabular}{lcc|cccccc|cccccc}
\\
\hline
\hline
\noalign{\smallskip}
&  & &   & & \multicolumn{2}{c}{Transit = 1} &  &   & & &  \multicolumn{2}{c}{Transit = 3} & \\
\noalign{\smallskip}
\hlineB{3.5}
\noalign{\smallskip}
Species & Bands & Wavelength  &  TNG & CAHA & Gemini & Keck & TMT & E-ELT &  TNG & CAHA & Gemini & Keck & TMT & E-ELT \\
 &  [$\mu$m] & range [$\mu$m]&  & & & & & & & & &  \\
\noalign{\smallskip}
\hlineB{3.5}
\noalign{\smallskip}
$\mathrm{H_2O}$ & 0.90 & 0.89 - 0.99 & 3.0 [ND] & 1.7 [ND] & -- & 8.1 [4.3] & 15.8 [9.4] & 16.6 [11.7] & 6.3 [3.8] & 6.2 [2.7] & -- & 14.2 [6.4] &  16.8 [13.8] & 18.3 [16.1] \\
& 0.94 & 0.93 - 0.95 & -- & 1.7 [ND] & -- & --  &  -- & 10.6 [8.4] & -- & 3.7 [1.7] & -- & -- &  -- & 12.6 [10.6] \\
& 1.00 & 0.95 - 1.05 & 2.7 [ND] & 2.2 [ND] & -- & 8.8 [3.6] &  15.9 [9.4] & 16.1 [12.0] & 5.9 [2.9] & 3.6 [1.2] & -- & 11.6 [9.0] &  16.2 [11.2] & 17.7 [15.3] \\
& 1.10 & 1.1 - 1.2 & 3.6 [ND] & 1.6 [ND] & -- & 3.9 [ND] &  4.5 [2.7] & 11.2 [6.4] & 3.7 [1.7] & 2.2 [1.2] & -- & 4.3 [2.0] &  5.5 [5.0] & 13.4 [7.7] \\
& 1.30 & 1.3 - 1.55 & ND & ND & 4.3 [1.6] & ND  &  ND & ND & ND & ND & 6.1 [4.2] & ND & ND & ND \\
\noalign{\smallskip}
\hline
\noalign{\smallskip}
$\mathrm{CH_4}$ & 0.89 & 0.875 - 0.91 & -- & ND & -- & --  &  -- & 6.4 [3.0] & -- & ND & -- & --  &  -- & 11.0 [5.6] \\
& 1.00 & 0.98 - 1.05 & 1.8 [ND] & 1.7 [ND] & -- & 3.8 [ND] &  8.3 [1.9] & 8.5 [2.1] & 3.9 [ND] & 2.4 [ND] & -- & 5.3 [1.1] &  10.4 [2.4] & 12.4 [2.9] \\
& 1.10 & 1.1 - 1.2 & ND & ND & -- & 1.6 [ND] &  2.3 [ND] & 5.2 [1.4] & 3.3 [1.3] & 1.4 [ND]& -- & 3.4 [1.7] &  3.7 [2.9] & 8.3 [4.8] \\
& 1.30 & 1.3 - 1.55 & ND & ND & 1.3 [ND] & ND  &  ND & ND & ND & ND & 2.4 [ND] & ND  &  ND & ND \\
& 1.60 & 1.6 - 1.8 & 5.9 [4.0] & 4.8 [2.1] &  6.8 [6.0] & 10.1 [6.1]  &  12.1 [11.9] & 15.5 [13.2] & 6.1 [5.8] & 5.7 [3.7] & 11.0 [9.6] & 11.2 [9.8] &  18.5 [17.2] & 19.0 [17.4] \\
& 2.40 & 2.2 - 2.5 & ND & -- & ND &  4.4 [2.1] &  17.3 [16.6] & -- & ND & -- & ND & 7.2 [6.2] &  20.1 [19.7] & -- \\
\noalign{\smallskip}
\hline
\noalign{\smallskip}
$\mathrm{C_2H_2}$ & 1.55 & 1.50 - 1.59 & ND & ND & ND & ND  &  ND & 2.2 [1.6] & ND & ND & ND & ND  &  2.0 [ND] & 2.4 [2.2] \\
\noalign{\smallskip}
\hline
\noalign{\smallskip}
CO & 1.55 & 1.555 - 1.60 &  ND & ND & 2.0 [ND] & 2.5 [1.2] &   4.2 [2.2] & 5.2 [3.4] & 1.9 [ND] & 1.4 [ND]& 2.2 [ND] & 3.2 [1.9] &  5.9 [3.7] & 6.4 [4.0] \\
& 2.30 & 2.3 - 2.5 & ND & -- & ND & 1.6 [ND] &  2.1 [2.0] & -- & ND & -- & ND & 2.7 [1.6] &  3.1 [2.8] & -- \\
\noalign{\smallskip}
\hline
\noalign{\smallskip}
$\mathrm{CO_2}$ & 1.59 & 1.45 - 1.66 &  ND & ND & ND & ND  &   ND & ND & ND & ND & ND & ND &  1.2 [ND] & 2.6 [ND] \\
\noalign{\smallskip}
\hline
\noalign{\smallskip}
HCN & all bands &  &  ND & ND & ND & ND  &   ND & ND & ND & ND & ND & ND  &  ND & ND \\
\noalign{\smallskip}
\hline
\noalign{\smallskip}
$\mathrm{NH_3}$ & 0.90 & 0.875 - 0.925 & -- & ND & -- & --  &  -- & 1.9 [ND] & -- & 2.3 [ND] & -- & --  &  -- & 3.3 [1.5] \\
& 1.00 & 1.0 - 1.1 & 2.0 [ND] & ND & -- & 3.6 [ND] &  10.4 [2.7] & 12.6 [3.8] & 2.6 [ND] & 2.3 [ND] & -- & 6.5 [1.9] &  13.3 [3.7] & 14.7 [4.0] \\
& 1.50 & 1.45 - 1.65 & ND & ND & ND & ND  &  3.8 [ND] & 6.7 [2.2] & ND & ND & ND & 1.4 [ND] &  6.8 [2.2] & 8.4 [3.8] \\
& 2.00 & 2.0 - 2.1  & ND & -- & ND & ND  &  3.4 [ND] & -- & ND & -- & ND & ND  &  4.0 [ND] & -- \\
& 2.30 & 2.2 - 2.4 & ND & -- & 3.4 [3.1] & 3.4 [3.1] &  8.6 [8.5] & -- & ND & -- & 5.8 [5.2] & 5.7 [5.1] &  9.7 [9.3] & -- \\
\noalign{\smallskip}
\hline
\noalign{\smallskip}
SH & all bands &  & ND & ND & ND & ND  &  ND & ND & ND & ND & ND & ND  &  ND & ND \\\noalign{\smallskip}
\hline
\noalign{\smallskip}
SN & all bands &  & ND & ND & ND & ND  &  ND & ND & ND & ND & ND & ND &  ND & ND \\
\noalign{\smallskip}
\hline
\noalign{\smallskip}
SO & all bands &  & ND & ND & ND & ND  &  ND & ND & ND & ND & ND &  ND &  ND & ND \\
\noalign{\smallskip}
\hline
\noalign{\smallskip}
$\mathrm{H_2S}$ & 1.60 & 1.55 - 1.66 & 2.1 [ND] & 1.2 [ND] & 4.2 [2.1] & 4.3 [2.2] &  7.0 [3.3] & 8.9 [5.4] & 2.8 [ND] & 1.7 [ND] & 6.6 [4.4] & 6.6 [4.1] &  9.7 [7.1] & 11.4 [7.8] \\
\noalign{\smallskip}
\hline
\hline
\label{tab:detetcion_k225b}
\end{tabular}}
\end{minipage}
\end{sidewaystable*}



\begin{sidewaystable*}
    
\vspace{-9cm}
\rotatebox{180}{
           \begin{minipage}{1.05\textwidth}
\centering
\caption{Similar to Tables \ref{tab:detetcion_hatp11b}, \ref{tab:detetcion_hd63433c}, and \ref{tab:detetcion_k225b} but for TOI-270 d.}
\resizebox{\textwidth}{!}{
\begin{tabular}{lcc|cccccc|cccccc}
\\
\hline
\hline
\noalign{\smallskip}
&  & &   & & \multicolumn{2}{c}{Transit = 1} &  &   & & &  \multicolumn{2}{c}{Transit = 3} & \\
\noalign{\smallskip}
\hlineB{3.5}
\noalign{\smallskip}
Species & Bands & Wavelength  &  TNG & CAHA & Gemini & Keck & TMT & E-ELT &  TNG & CAHA & Gemini & Keck & TMT & E-ELT \\
 &  [$\mu$m] & range [$\mu$m]&  & & & & & & & & &  \\
\noalign{\smallskip}
\hlineB{3.5}
\noalign{\smallskip}
$\mathrm{H_2O}$ & 0.90 & 0.89 - 0.99 & 6.7 [4.9] & 5.1 [3.1] & -- & 12.8 [10.4] &  17.7 [16.9] & 19.8 [18.4] & 9.3 [8.0] & 7.1 [6.1] & -- & 16.4 [14.6] &  18.3 [18.1] & 20.9 [19.5] \\
& 0.94 & 0.93 - 0.95 & -- & 3.4 [1.8] & -- & --  &  -- & 14.2 [13.8] & -- & 4.5 [2.4] & -- & --  &  -- & 16.7 [15.9] \\
& 1.00 & 0.95 - 1.05 & 6.5 [5.4] & 3.6 [2.9] & -- & 14.7 [8.6]  &  16.5 [16.1] & 18.8 [17.1] & 9.6 [8.4] & 7.1 [6.4] & -- &  17.2 [14.4] &  18.0 [17.5] & 20.4 [18.9] \\
& 1.10 & 1.1 - 1.2 & 4.9 [3.9] & 2.2 [ND] & -- &  5.5 [4.3] &  8.1 [6.0] & 11.8 [7.7] & 7.7 [6.1] & 3.7 [2.1] & -- & 9.2 [6.8]  &  13.0 [10.0] & 15.9 [15.2] \\
& 1.30 & 1.3 - 1.55 & ND & ND & 5.9 [5.5] & ND  &  ND & ND & ND & ND & 8.8 [7.3] & ND  &  ND & ND \\
\noalign{\smallskip}
\hline
\noalign{\smallskip}
$\mathrm{CH_4}$ & 0.89 & 0.875 - 0.91 & -- & 1.2 [ND] & -- &  -- &  -- & 8.9 [8.1] & -- & 2.1 [1.6] & -- & --  &  -- & 12 [9.8] \\
& 1.00 & 0.98 - 1.05 & ND & ND  & -- & 2.5 [1.9]  &  9.3 [4.1] & 10.5 [4.4] & 1.7 [ND] & 1.4 [ND] & -- & 6.2 [2.1]  &  12.1 [5.6] & 13.5 [7.9] \\
& 1.10 & 1.1 - 1.2 & ND & ND & -- &  1.7 [ND] &  2.5 [2.3] & 5.1 [4.3] & ND & ND & -- &  1.9 [1.7] &  5.3 [2.8] & 12.0 [6.2] \\
& 1.30 & 1.3 - 1.55 & ND & ND & 3.3 [ND] & ND  &  ND & ND & ND & ND & 3.9 [1.9] & ND  &  ND & ND \\
& 1.60 & 1.6 - 1.8 & 7.9 [6.4] & 5.0 [4.5] & 13.2 [10.9] & 14.2 [11.5]  &  17.8 [15.2] & 20.4 [16.4] & 9.1 [8.5] & 8.3 [7.7] & 14.7 [13.6] & 15.5 [14.1]  &  19.6 [16.7] & 21.1 [17.1] \\
& 2.40 & 2.2 - 2.5 & ND & -- & 3.1 [1.3] & 13.7 [11.4]  &  20.5 [19.3] & -- & 4.2 [2.0] & -- &  5.4 [4.4] & 14.9 [14.2]  &  22.8 [21.6] & -- \\
\noalign{\smallskip}
\hline
\noalign{\smallskip}
CO & 1.55 & 1.555 - 1.60 &  2.4 [ND] & 1.7 [ND] & 5.9 [4.8] & 6.2 [5.3]  &   10.8 [10.2] & 12.2 [10.6] & 4.3 [3.3] & 2.1 [1.5] & 7.4 [6.2] & 8.8 [7.9]  &  12.9 [11.1] & 14.1 [13.4] \\
& 2.30 & 2.3 - 2.5 & ND & -- & ND & 1.2 [ND]  &  4.5 [3.5] & -- & ND & -- & ND & 3.6 [2.1]  &  6.1 [5.3] & -- \\
\noalign{\smallskip}
\hline
\noalign{\smallskip}
$\mathrm{CO_2}$ & 1.59 & 1.45 - 1.66 &  ND & ND & ND & ND  &   ND & 3.0 [2.0] & ND & ND & ND & ND  &  ND & 6.8 [5.9] \\
& 2.00 & 1.94 - 2.08 & ND & -- & 4.2 [2.6] & 4.5 [3.3]  &  6.9 [3.3] & -- & ND & -- & 5.7 [3.7] & 5.8 [3.9]  &  8.0 [4.7] & -- \\
\noalign{\smallskip}
\hline
\noalign{\smallskip}
HCN & all bands &  &  ND & ND & ND & ND  &   ND & ND & ND & ND & ND & ND  &  ND & ND \\
\noalign{\smallskip}
\hline
\noalign{\smallskip}
$\mathrm{NH_3}$ & 0.90 & 0.875 - 0.925 & -- & ND & -- & --  &  -- & 4.3 [2.3] & -- & ND & -- & --  &  -- & 4.7 [3.7] \\
& 1.00 & 1.0 - 1.1 & 2.7 [2.1] & 1.9 [1.5] & -- & 6.7 [2.7]  &  11.2 [6.7] & 13.6 [9.6] & 4.7 [2.8] & 3.1 [2.5] & -- & 7.6 [3.2]  &  14.7 [9.6] & 16.1 [12.9] \\
& 1.50 & 1.45 - 1.65 & 3.2 [1.6] & ND & 4.0 [3.7] & 4.3 [3.5]  &  4.6 [3.8] & 8.1 [5.9] & 4.3 [3.5] & 1.5 [ND] & 5.1 [4.6] & 5.3 [4.5]  &  6.6 [5.8] & 10.0 [8.8] \\
& 2.00 & 2.0 - 2.1  & ND & -- & 1.3 [ND] & 1.3 [ND]  &  2.5 [ND] & -- & 3.7 [2.5] & -- & 5.1 [3.7] & 4.6 [3.1]  &  5.6 [5.0] & -- \\
& 2.30 & 2.2 - 2.4 & 2.9 [ND] & -- & 7.6 [5.8] & 7.7 [6.0]  &  11.9 [10.4] & -- & 4.1 [3.7] & -- & 8.8 [7.9] & 9.1 [8.4]  &  14.8 [13.7] & -- \\
\noalign{\smallskip}
\hline
\noalign{\smallskip}
SO & all bands &  & ND & ND & ND & ND  &  ND & ND & ND & ND & ND &  ND &  ND & ND \\
\noalign{\smallskip}
\hline
\noalign{\smallskip}
SCO & all bands &  & ND & ND & ND & ND  &  ND & ND & ND & ND & ND & ND  &  ND & ND \\
\noalign{\smallskip}
\hline
\noalign{\smallskip}
$\mathrm{CS_2}$ & 1.94 & 1.92 - 1.96 & ND & -- & ND & ND  &  2.9 [ND] & -- & ND & -- & ND & ND  &  3.1 [1.5] & -- \\
& 2.23 & 2.14 - 2.32 & ND & -- & ND & ND  &  ND & -- & ND & -- & 1.2 [ND] &  ND &  2.0 [1.9] & -- \\
\noalign{\smallskip}
\hline
\noalign{\smallskip}
$\mathrm{H_2S}$ & 1.00 & 1.0 - 1.03 & ND & ND & -- &  1.2 [ND] &  2.7 [1.9] & 3.0 [2.8] & ND  & ND & -- & 2.1 [1.5]  &  4.0 [2.3] & 6.5 [3.5] \\
& 1.60 & 1.55 - 1.66 & 6.7 [5.6] & 4.0 [2.8] & 10.1 [7.1] &  11.1 [6.2] &  14.8 [13.6] & 17.4 [15.5] & 7.0 [6.0] & 6.6 [5.3] & 13.2 [11.7] &  13.9 [10.5] &  15.9 [15.2] & 18.9 [16.6] \\
\noalign{\smallskip}
\hline
\hline
\label{tab:detetcion_toi270d}
\end{tabular}}
\end{minipage}}
\end{sidewaystable*}
\begin{figure*}
    \centering  
        \begin{minipage}[b]{1.9\columnwidth}
            \includegraphics[width=\columnwidth]{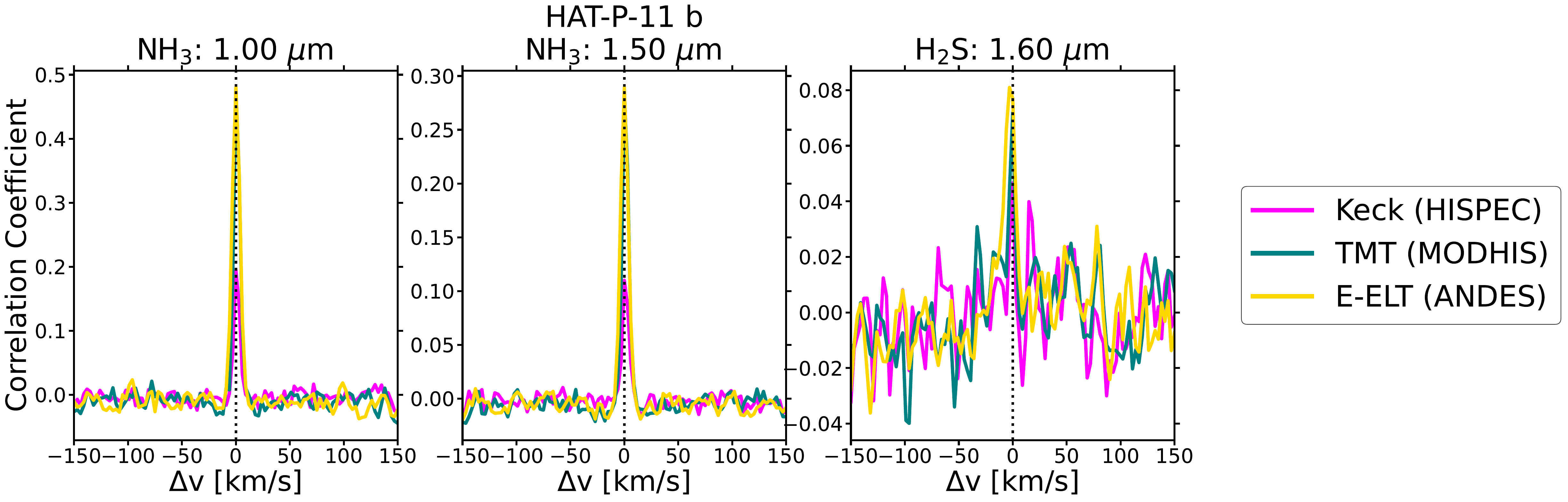}
        \end{minipage}
         \begin{minipage}[b]{2\columnwidth}
            \includegraphics[width=\columnwidth]{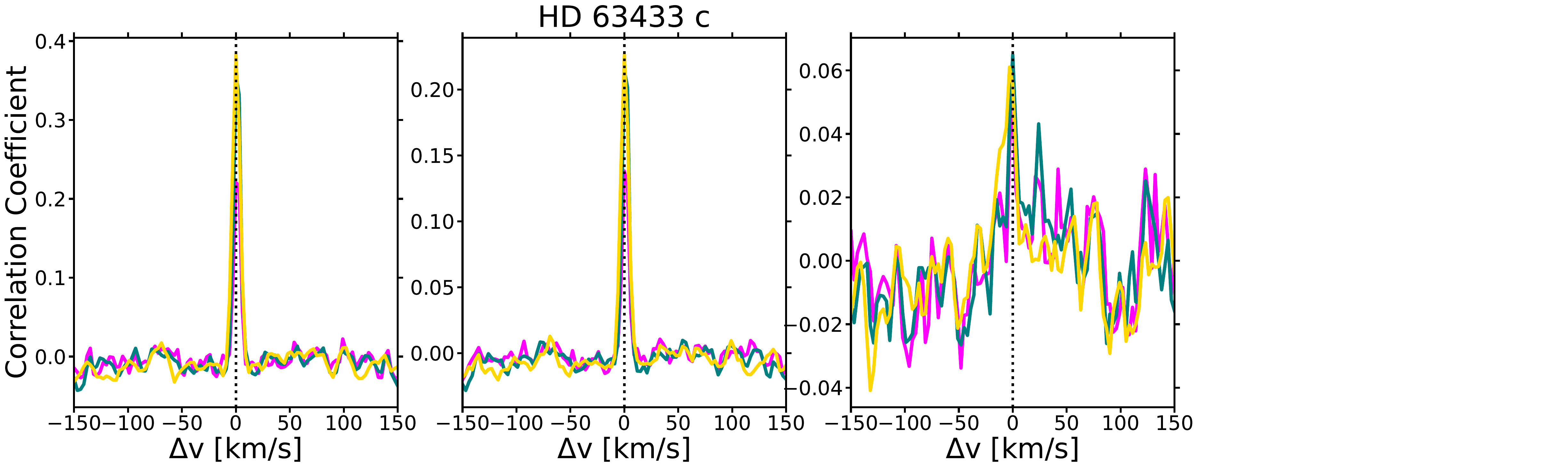}
        \end{minipage}
         \begin{minipage}[b]{2\columnwidth}
            \includegraphics[width=\columnwidth]{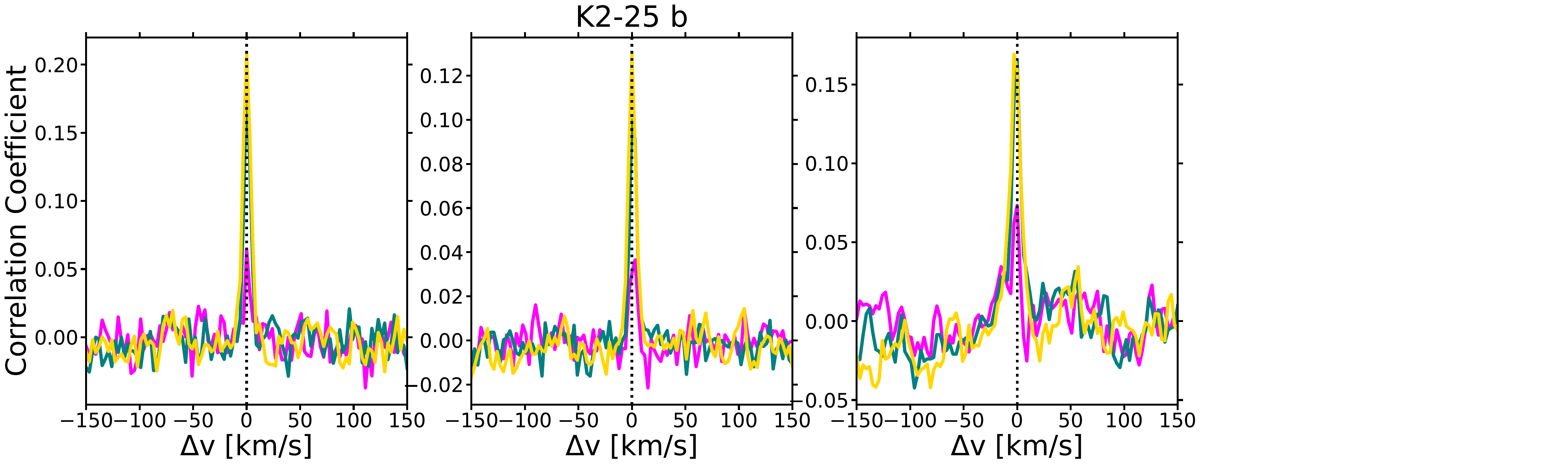}
        \end{minipage}
        \begin{minipage}[b]{2\columnwidth}
            \includegraphics[width=\columnwidth]{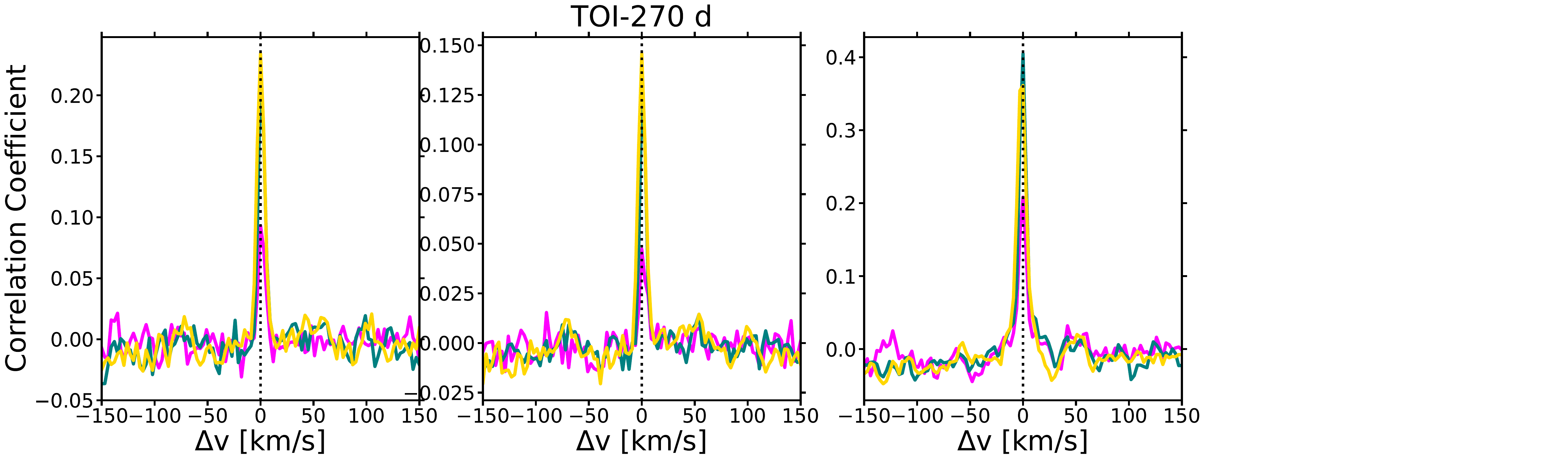}
        \end{minipage}


\caption{Correlation coefficient vs. velocity shift map for the 1.00 and 1.50 $\mathrm{\mu m}$ bands of $\mathrm{NH_3}$ and the 1.60 $\mathrm{\mu m}$ band of $\mathrm{H_2S}$ across the atmospheres of Neptune-class planets, using future high-resolution spectrographs (Keck: \textit{magenta}, TMT: \textit{teal}, and E-ELT: \textit{gold}) over three transits in cloud-free conditions. Each row corresponds to a specific planet (\textit{from top to bottom}: HAT-P-11 b, HD 63433 c, K2-25 b, and TOI-270 d). These bands are detectable by all three instruments with enhanced sensitivity, highlighting the potential of future ground-based instruments to constrain nitrogen and sulfur chemistry.}

\label{fig:CCF_neptune}
\end{figure*}

\section{Results}
\label{sec:results}

\subsection{Molecular chemistry and variability across different classes of Neptune atmospheres}
\label{result:chemistry}

The various classes of Neptune-like atmospheres exhibit diverse and complex chemistry, significantly influenced by their unique physical and atmospheric characteristics as well as the varying levels of stellar UV radiation they receive. In this section, we discuss the dominant chemical processes in these atmospheres and how various mechanisms influence the formation and destruction of molecules (see Appendix \ref{appendix:chemistry} for details). The analysis focuses on molecules that are significantly detectable using various spectrographs in a clear atmosphere, with a $\mathrm{\sigma_{det}}$ $>$ 3 across 3 transits (refer to Figure \ref{fig:summary} for a summary derived from the investigation detailed in Section \ref{result:cross-correlation}). Compared to hot and ultra-hot Jupiters, these planets have lower $T_{\mathrm{eq}}$. Additionally, the presence of a strong vertical mixing leads to the quenching of the most abundant molecules across nearly all scenarios. The VMRs for the most abundant species in the atmospheres are illustrated in Figure \ref{fig:VMR} and the resultant transmission spectra for each planet are shown in Figure \ref{fig:Transmission_spectra} with respective contributions from different molecules.

For HAT-P-11 b, quenching is less effective due to its relatively higher atmospheric temperature structure, which can partially compete with the effects of vertical mixing. This higher temperature facilitates molecular diffusion in each atmospheric layer. HD 63433 c and K2-25 b experience quenching of $\mathrm{H_2O}$, $\mathrm{CH_4}$, and CO, with K2-25 b also exhibiting quenching of $\mathrm{NH_3}$ due to its lower temperature compared to HD 63433 c. TOI-270 d displays an almost completely quenched atmosphere for all abundant molecules, attributed to its significantly lower atmospheric temperature. As noted by \cite{gao2021aerosols}, HAT-P-11 b's temperature is suitable to have a cloud-free atmosphere, resulting in feature-rich spectra. The chemical analysis indicates that the atmosphere does not produce higher hydrocarbons, unlike the other planets. While TOI-270 d does produce $\mathrm{CH_3OH}$ and $\mathrm{H_2CO}$, the average concentrations in the photosphere level are not significant ($\mathrm{10^{-9}}$ $<$ VMR $<$ $\mathrm{10^{-8}}$). Same goes for HD 63433 c (with $\mathrm{C_2H_4}$, $\mathrm{C_2H_6}$, $\mathrm{C_3H_4}$, $\mathrm{C_4H_3}$, and $\mathrm{C_6H_6}$ in the atmosphere) and K2-25 b (with $\mathrm{C_2H_6}$ and $\mathrm{C_3H_4}$ in the atmosphere). Here, the discussion focuses on the potential for having hydrocarbon hazes in these atmospheres. The abundances of $\mathrm{CH_4}$, $\mathrm{NH_3}$, and $\mathrm{H_2S}$ in the atmosphere of HAT-P-11 b are relatively higher because these molecules rapidly recombine after being destroyed to their precursors. The formation and destruction of $\mathrm{CH_4}$ and $\mathrm{H_2S}$ in the atmosphere of HAT-P-11 b are primarily driven by molecular diffusion. For $\mathrm{NH_3}$, the processes are also similar except for the photochemical destruction of $\mathrm{NH_3}$ to $\mathrm{NH_2}$, which accounts for nearly 38\% of the total $\mathrm{NH_3}$ destruction. The lower abundance of $\mathrm{H_2O}$ is associated with a supersolar C/O ratio.

Conversely, HD 63433 c is dominated by $\mathrm{H_2O}$, $\mathrm{CH_4}$, CO, $\mathrm{CO_2}$, HCN, $\mathrm{NH_3}$, and $\mathrm{H_2S}$ along with a complex hydrocarbon $\mathrm{C_2H_2}$. While all molecules are dominated by the molecular diffusion process, CO is present in the atmosphere as a photochemical product (photochemical dissociation of HCO contributes to the formation of 28\% of the total CO in the atmosphere). Similar to HAT-P-11 b, faster recombination processes are responsible for higher abundance of all molecules. Among all, $\mathrm{H_2O}$, $\mathrm{CO_2}$, and $\mathrm{NH_3}$ undergo photochemical destruction to their respective sink molecules: $\mathrm{H_2O}$ to OH (100\%), $\mathrm{CO_2}$ to CO (36\%), and $\mathrm{NH_3}$ to $\mathrm{NH_2}$ (100\%).

The atmosphere of K2-25 b is primarily composed of $\mathrm{H_2O}$, $\mathrm{CH_4}$, CO, $\mathrm{NH_3}$, and $\mathrm{H_2S}$. As observed for the previous two cases, elevated levels of these gases are similarly due to accelerated formation processes. Similar to HAT-P-11 b, the majority of molecules are formed through dominant molecular diffusion pathways, with the exception of $\mathrm{NH_3}$, which undergoes photodissociation to $\mathrm{NH_2}$, accounting for 27\% of the total $\mathrm{NH_3}$ destruction.

Lastly, the atmosphere of TOI-270 d was recently observed with JWST and found to contain a variety of molecules, suggesting a near-solar C/O ratio and a metal-rich composition. Consistent with the findings of \cite{benneke2024jwst}, we found significant presence of $\mathrm{H_2O}$, $\mathrm{CH_4}$, $\mathrm{CO_2}$, and tentatively, $\mathrm{CS_2}$. Additionally, CO (a photochemical product of SCO), $\mathrm{NH_3}$, and $\mathrm{H_2S}$ were also found to be other dominating species in the atmosphere. The primary destruction processes for $\mathrm{CH_4}$, CO, $\mathrm{CO_2}$, and $\mathrm{H_2S}$ are dominated by two-body molecular diffusion reactions. Additionally, a notable fraction of $\mathrm{H_2S}$ (27\%) undergoes photodissociation to form SH. Photochemical processes are also significant in the destruction of $\mathrm{H_2O}$ to OH (100\%), $\mathrm{NH_3}$ to $\mathrm{NH_2}$ (100\%), and $\mathrm{CS_2}$ to CS (100\%). However, rapid recombination of these product molecules contributes to maintaining higher concentrations of the original abundant species.


\subsection{Estimating the detectability of molecules from the synthetic observations using cross-correlation spectroscopy}
\label{result:cross-correlation}

As previously discussed in \cite{currie2023there} and \cite{2024ApJ...972..165D}, SPECTR deviates from the traditional S/N estimation, where an S/N of zero indicates a non-detection. Instead, given the complex nature of spectral lines and their interactions, SPECTR utilizes a more rigorous detection scheme, which interprets a non-detection as having a significance level of $\mathrm{\sigma_{det}} \sim$ 1.  Instead of pursuing molecular detection across the entire wavelength range, we focused on specific spectral bands for each molecule (see Table \ref{tab:detetcion_hatp11b}, \ref{tab:detetcion_hd63433c}, \ref{tab:detetcion_k225b}, and \ref{tab:detetcion_toi270d}) and analyzed their detectability over multiple transits. This targeted approach allows us to identify the particular bands to examine when attempting to detect molecules in various classes of Neptune atmospheres. Additionally, considering the presence of clouds using a cloudy model (with a gray cloud deck at 0.01 bar and a haze layer) provides a comparative understanding of how Rayleigh scattering influence the detectability measures of gaseous species in the atmosphere. This helps determine the optimal spectral windows for different molecules and spectrographs, considering the impact of clouds on detection efficiency. In the following sections, we provide an in-depth analysis of molecular detection in cloudy and cloud-free atmospheres for all four planets, utilizing the six spectrographs of interest and identifying the optimal conditions and instruments for observing different planetary atmospheres. We summarize the detectability based on three transits of the planets to present the best-case scenario for the instruments. 

HAT-P-11 b is the warmest planet in our study. Its hot, extended atmosphere and the hypothesized absence of clouds make it an excellent candidate for investigating Neptune-class atmospheres. A summary of molecular detectability is listed in Table \ref{tab:detetcion_hatp11b}. \textbf{Despite the detection of $\mathrm{H_2O}$ by \cite{fraine2014water} and \cite{basilicata2024gaps}, our results fail to detect $\mathrm{H_2O}$ on HAT-P-11 b. This is attributed to the atmospheric chemistry associated with a super-solar C/O ratio (see Section \ref{sec:atmosphere}); hence, the planet does not produce a significant amount of $\mathrm{H_2O}$ in its atmosphere. HD 63433 c, a warm sub-Neptune class planet, is characterized by distinct chemistry and geometry compared to other Neptune-class planets. Its $T_{\mathrm{eq}}$ is highly sensitive to the presence or absence of clouds. Therefore, it offers a comprehensive view of molecular detectability through a complementary analysis of both clear and cloudy atmospheres (see Table \ref{tab:detetcion_hd63433c}). The atmospheres of K2-25 b and TOI-270 d are rich in molecules, and cloud cover has minimal impact on the visibility of molecular features (see Tables \ref{tab:detetcion_k225b} and \ref{tab:detetcion_toi270d}). The recent JWST observation of TOI-270 d by \cite{benneke2024jwst} highlights the planet as a compelling target for studying its underlying atmospheric chemistry and comparing the detectability of molecules using ground-based observatories. For HD 63433 c, K2-25 b, and TOI-270 d, IGRINS demonstrates superior performance over other instruments in detecting the 1.30 $\mu$m band of $\mathrm{H_2O}$. Other instruments fail to detect this particular $\mathrm{H_2O}$ band due to the interference from Earth's atmospheric telluric lines near 1.30 $\mu$m. IGRINS surpasses these limitations as its spectral coverage starts from 1.45 $\mu$m, thus experiencing less superimposition with telluric noise in this region.}






\section{Conclusions and Summary}
\label{sec:conclusion}
With recent advancements in high-resolution observations of hot and ultra-hot Jupiter atmospheres, the ongoing quest of exploring Neptune-class planets using the cross-correlation technique has become essential for several reasons. As cooler and smaller planets compared to hotter gas giants, studying these atmospheres improves our current understanding of disequilibrium chemistry and how they influence the atmospheric compositions of Neptunian planets \citep{guilluy2022gaps}.

\begin{figure*}
\centering
	\includegraphics[width=\textwidth]{Figures/summary.pdf}
  \caption{Summarization of molecular detectability (with $>$ 3$\mathrm{\sigma_{det}}$) over three transits for four different subclasses of planets, using recent and upcoming ground-based spectrographs. The influence of cloudy atmospheres on molecular features and their detectability is highlighted. Additionally, the formation processes of the molecules at a specific pressure layer of 0.01 bar are discussed. Planetary visualizations are sourced from the \href{https://science.nasa.gov/exoplanets/exoplanet-catalog/}{NASA Exoplanet Catalog}.}
   \label{fig:summary}
\end{figure*}

In this study, we investigate four classes of Neptune atmospheres (\textit{warm Neptune}: HAT-P-11 b, \textit{warm sub-Neptune}: HD 63433 c, \textit{temperate Neptune}: K2-25 b, and \textit{temperate sub-Neptune}: TOI-270 d) for one and three transits using current and future ground-based spectrographs of different resolutions, collecting areas, and complementary wavelength coverages: GIANO-B, CARMENES, IGRINS, HISPEC, MODHIS, and ANDES. These classes of planets possess variability in their PT structures, resulting in distinct atmospheric chemical processes. Specifically, we aim to illustrate how upcoming instruments such as HISPEC, MODHIS, and ANDES can surpass current ground-based observatories (GIANO-B, CARMENES, and IGRINS) in detecting multiple molecules in both clear and cloudy atmospheres. In addition, the three upcoming instruments have the potential to detect nitrogen- and sulfur-bearing species in Neptune-like atmospheres, both of which are poorly understood, shedding light on the significance of nitrogen and sulfur chemistry occurring within these atmospheres (refer to Figure \ref{fig:CCF_neptune}, which presents the correlation coefficient vs. velocity shift map for specific $\mathrm{NH_3}$ and $\mathrm{H_2S}$ bands using HISPEC, MODHIS, and ANDES over three transits under cloud-free conditions). A brief overview of the detectability of different molecules is shown in Figure \ref{fig:summary}.

\begin{itemize}
    
    \item ANDES demonstrates superior sensitivity in detecting molecular bands with higher detection significance compared to other spectrographs. This advantage stems from its ability to probe atmospheres with higher resolution and larger collecting area. \textbf{While MODHIS approaches the performance level of ANDES, it demonstrates slightly lower detection significances compared to ANDES for common wavelength bands. This difference is primarily due to its smaller collecting area compared to ANDES. A reduced collecting area limits the instrument's ability to gather sufficient photons, thereby decreasing the S/N and reducing sensitivity to molecular features.} Nevertheless, simultaneous observations with ANDES and MODHIS could unlock new possibilities for characterizing four classes of Neptune atmospheres by leveraging their combined ability to probe different spectral bands. The overall performance follows the order: ANDES $>$ MODHIS $>$ HISPEC $>$ IGRINS $>$ GIANO-B $>$ CARMENES. HISPEC performs slightly better than IGRINS in molecular detection due to higher resolution and collecting area. GIANO-B outperforms CARMENES primarily due to lower read noise, reduced thermal noise from the instrument, and a slightly larger collecting area.
   
    \item Clouds have a substantial impact on planetary spectra and hence, the detectability of molecular fingerprints. They can obscure molecular features in optical bands but exert less influence beyond 1.6 $\mu$m. Therefore, investigating near-infrared bands can offer more accurate observations and constraints on atmospheric composition. As mentioned previously, HISPEC, MODHIS, and ANDES are less affected by clouds due to their superior resolution, which allows molecular line cores to extend well above the cloud deck, thereby enhancing the ability to probe atmospheric components.
    
    \item HAT-P-11 b: Previous observations from both space and ground have detected $\mathrm{H_2O}$ in the atmosphere of the planet \citep{fraine2014water,basilicata2024gaps}. However, we have failed to detect water on HAT-P-11 b, likely due to its lower atmospheric abundance, which is related to our assumption of a super-solar C/O ratio. Combined observations from \textit{HST} WFC3, STIS, and \textit{Spitzer} have yielded a range of retrieved C/O ratios for HAT-P-11 b's atmosphere, varying from 1.03 (from \textit{HST} WFC3) to 0.63 (from \textit{HST} WFC3 + \textit{Spitzer}). We have adopted the value derived from \textit{HST} WFC3 + STIS observations due to the consistency between the \textit{HST} WFC3 and \textit{HST} WFC3 + STIS data. \textbf{Figure \ref{fig:trans_high_low_hatp11b} illustrates a comparison between the observed data from the \textit{HST} WFC3 + STIS, as reported in \cite{chachan2019hubble}, and our high-resolution model spectra (R=50000). The data clearly indicate the presence of $\mathrm{H_2O}$ in the planet's atmosphere, with prominent water absorption features observed in the 1.1–1.2 $\mathrm{\mu}$m and 1.3–1.5 $\mathrm{\mu}$m regions. As discussed earlier, our model spectra fall short of reproducing the observed water signals, primarily due to the higher retrieved C/O ratio from \cite{chachan2019hubble}, which was incorporated into our chemical models. While the retrievals from the \textit{HST} WFC3 + STIS data allow for a wide range of C/O ratios, with a 68\% confidence interval spanning C/O = 0.51 to C/O = 1.56, the median value is approximately C/O $\approx$ 1. This substantial range suggests that the atmospheric composition could transition from being $\mathrm{H_2O}$-dominated to $\mathrm{CH_4}$-dominated as the C/O ratio increases. These findings underscore the importance of applying an improved retrieval framework to the data to refine molecular abundance estimates and achieve tighter constraints on the C/O ratio. To further assess the detectability of $\mathrm{H_2O}$ in the planet's atmosphere, future \textit{JWST} observations will be helpful, as they will provide more precise constraints on the C/O ratio.} GIANO-B observations identified $\mathrm{NH_3}$ and $\mathrm{CH_4}$ in the atmosphere with 5$\mathrm{\sigma}$ and 2.6$\mathrm{\sigma}$ significance, respectively, over four transit observations. Our results align with detections in a cloudy atmosphere, albeit with slightly different detectability for three transits: 5.7$\mathrm{\sigma_{det}}$ for the 1.50 $\mu$m band of $\mathrm{NH_3}$ band and 5.4$\mathrm{\sigma_{det}}$ for the 1.60 $\mu$m band of $\mathrm{CH_4}$. This suggests the potential presence of clouds in HAT-P-11 b's atmosphere. A more detailed study of cloud modeling could reveal the possible microphysics on the planet. Additionally, IGRINS, HISPEC, MODHIS, and ANDES can detect $\mathrm{H_2S}$ only in a cloud-free atmosphere with $>$ 3$\mathrm{\sigma_{det}}$.

    \begin{figure}
\centering
	\includegraphics[width=\columnwidth]{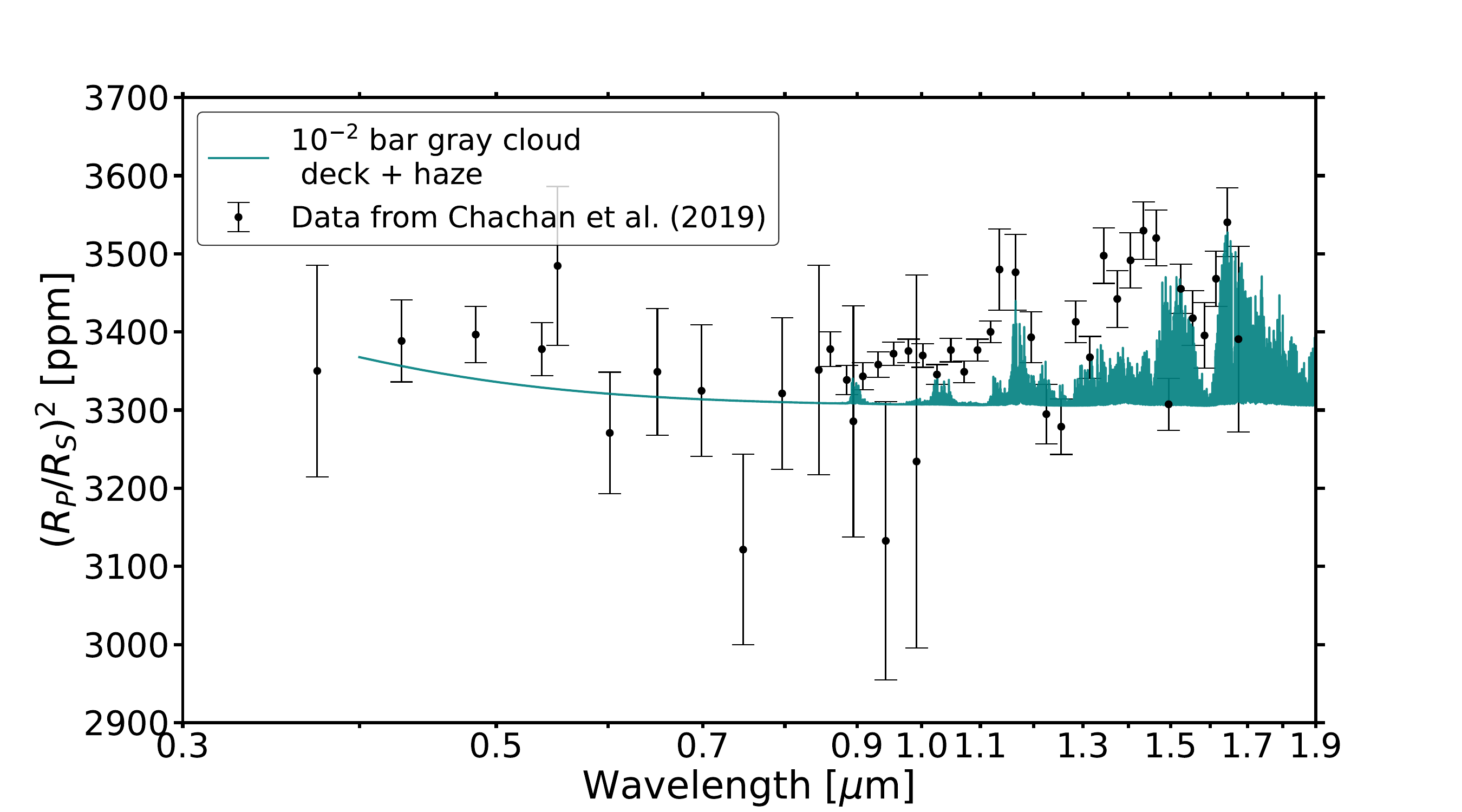}
  \caption{\textbf{Comparison of the high-resolution modeled spectra (R=50000) of HAT-P-11 b, derived from the retrieved parameters in \cite{chachan2019hubble}, with the lower-resolution observational data in the \textit{HST} WFC3 + STIS bandpass.}}
   \label{fig:trans_high_low_hatp11b}
\end{figure}
    
    \item HD 63433 c: $\mathrm{H_2O}$ and $\mathrm{CH_4}$ are detectable with all the instruments used. Ground-based observations can enhance our understanding of nitrogen chemistry by detecting HCN and $\mathrm{NH_3}$ in the atmosphere. This planet shows evidence of photochemical CO production. In a cloudy atmosphere, CO detection is possible only with IGRINS, HISPEC, MODHIS, and ANDES, requiring a minimum of three transits. In a cloud-free atmosphere, CO can be easily detected with just one transit observation with the same instruments except for three transits with GIANO-B. Similar to HAT-P-11 b, $\mathrm{H_2S}$ detection is feasible only in cloud-free conditions using IGRINS, HISPEC, MODHIS, and ANDES. ANDES can reveal the presence of the complex hydrocarbon $\mathrm{C_2H_2}$ in the planet. MODHIS can also detect $\mathrm{C_2H_2}$ albeit with lesser significance. Additionally, the detection of $\mathrm{CO_2}$ is only possible with MODHIS and it demands a cloud-free atmosphere.
    
    \item K2-25 b: GIANO-B and CARMENES can only detect $\mathrm{H_2O}$ and $\mathrm{CH_4}$, providing insights into carbon and oxygen chemistry. In contrast, IGRINS, HISPEC, MODHIS, and ANDES can probe $\mathrm{NH_3}$ and $\mathrm{H_2S}$, with HISPEC, MODHIS, and ANDES also capable of detecting CO. Similar to TOI-270 d, clouds have a lesser impact on this planet, as the same molecules are detectable in both cloud-free and cloudy atmospheres, except for $\mathrm{H_2O}$ with CARMENES. This is because $\mathrm{H_2O}$ bands are significantly affected by clouds, requiring superior instrumental architecture for detection.
    
    \item TOI-270 d: Similar to \cite{benneke2024jwst}, TOI-270 d's atmosphere can be characterized by $\mathrm{H_2O}$ and $\mathrm{CH_4}$ features using all spectrographs, even in the presence of clouds. However, detecting $\mathrm{CO_2}$ requires a larger collecting area, making it possible with IGRINS, HISPEC, MODHIS, and ANDES. \textbf{A larger collecting area increases the chance of molecular detection by improving the S/N, allowing fainter spectral features to be discerned from the noise. This is particularly crucial for $\mathrm{CO_2}$, as its spectral lines are weaker compared to other molecules like $\mathrm{H_2O}$. The detection of $\mathrm{CO_2}$ from ground-based spectroscopy also presents challenges due to its non-negligible abundance in Earth's atmosphere, which contaminates the observed spectra. Therefore, the successful detection of $\mathrm{CO_2}$ relies strongly on the accuracy of telluric correction methods and the advanced capabilities of instruments.} Similar to HD 63433 c, CO in TOI-270 d's atmosphere is a photochemical byproduct and indicates a high metal-rich atmosphere. $\mathrm{NH_3}$ and $\mathrm{H_2S}$ are detectable with all spectrographs, except CARMENES, which cannot substantially detect $\mathrm{NH_3}$. Additionally, the tentative detection ($>$ 3$\mathrm{\sigma_{det}}$) of the 1.94 $\mu$m band of $\mathrm{CS_2}$ in a clear atmosphere using MODHIS highlights the importance of sulfur chemistry in the atmosphere. This finding demands further investigation through detailed chemical studies and additional observations from both space and ground-based instruments.

    \item \textbf{SPECTR assumes ``perfect" telluric line removal, where the planetary signal remains unaffected after detrending.  While this approach is a reasonable starting point for assessing molecular detectability, it does not fully account for the challenges associated with telluric contamination in the case of slow-moving planets (with longer orbital period), such as TOI-270 d and HD 63433 c in our case, where the planetary radial velocity shift during transit is minimal. As highlighted by \cite{cheverall2024feasibility}, planetary signals may still be recoverable in such scenarios through PCA-based detrending, provided there are sufficient out-of-transit spectra. The contrast provided by the out-of-transit data prevents the planetary signal (present only during transit) from being absorbed into the principal components even in cases of subpixel radial velocity changes. However, the effectiveness of PCA-based detrending is highly sensitive to observational factors, such as the ratio of out-of-transit to in-transit spectra and the magnitude of the planetary pixel shift during transit. By assuming perfect telluric removal, this study actually provides an optimistic estimate of molecular detectability. For slow-moving planets, where the overlap between planetary and telluric signals is substantial, detection efficiencies are likely to be lower than predicted by the model. Future work could incorporate more realistic noise treatments, including imperfect detrending and observational constraints, to refine these predictions and provide a more accurate assessment of molecular detectability for planets with small radial velocity variations and long orbital periods.}
\end{itemize}

We further investigate the chemical pathways to identify the dominant processes contributing to the formation and destruction of various molecules (see Section \ref{result:chemistry}, \textbf{Appendix \ref{appendix:chemistry}} and Figure \ref{fig:summary} for details). In addition to molecular diffusion and atmospheric mixing, photochemistry plays a crucial role for some molecules. TOI-270 d, being the coldest planet in our study, experiences atmospheric quenching as Eddy diffusion prevails over chemical reactions (mixing timescale is smaller than chemical timescale). Future work could involve conducting ground-based observations of these planets with current spectrographs, analyzing the detection feasibility of the molecules discussed here, and linking the underlying chemical processes.

\textbf{This work focuses on the high-resolution transmission spectra of warm Neptune to temperate Sub-Neptune atmospheres using atmospheric forward models. Our aim is to investigate the role of disequilibrium chemistry and pressure-temperature profiles (PT profiles), which significantly influence the volume mixing ratio (VMR) profiles of molecules in these atmospheres. Additionally, we quantify the detectability of these molecules using ground-based cross-correlation spectroscopy. We acknowledge the importance of performing atmospheric retrievals on the high-resolution spectra generated by our forward models. Such retrievals are essential to statistically derive robust constraints on the VMRs and PT profiles of these atmospheres, enabling us to evaluate how accurately atmospheric parameters can be retrieved for a given observed spectrum. However, implementing high-resolution retrieval frameworks is computationally intensive and requires detailed, focused studies of individual planets in our sample, which is beyond the scope of this work (see, for example, \cite{blain2024formally}). Future studies focusing on these aspects will provide more detailed and statistically robust constraints for characterizing these atmospheres, particularly with upcoming extremely large telescopes equipped with high-resolution spectrographs.}

\section*{acknowledgements}
\noindent  L.M. acknowledges financial support from DAE and DST-SERB research grants [MTR/2021/000864] of the Government of India. D.D. and L.M. thank Dr. Paul Mollière for providing access to petitCODE and for actively participating in discussions regarding its use and applications in the past.  D.D. and L.M. thank Mr. Spandan Dash from the University of Warwick for valuable discussions regarding the observational aspects of high-resolution cross-correlation spectroscopy. D.D. and L.M. also thank Prof. Sergey Yurchenko from UCL for his suggestions on generating high-resolution opacities using the ExoCross package. D.D. and L.M. both thank Dr. Miles Currie from NASA GSFC for making the SPECTR simulator available to the community. This research was carried out, in part, at the Jet Propulsion Laboratory and the California Institute of Technology under a contract with the National Aeronautics and Space Administration. The high-resolution opacity files generated in this work will be made available on the petitRADTRANS website in the future. We would like to thank the anonymous referee for constructive comments that helped improve the manuscript.

\bibliographystyle{aasjournal}{}
\bibliography{references.bib}
\clearpage

\appendix
\section{Comparative analysis of chemistry and atmospheric processes across different classes of Neptune atmospheres}
\label{appendix:chemistry}

In this section, we present an overview of the chemical reactions driving the formation and destruction of various molecules within the 0.01 bar pressure layer across different classes of Neptune-like exoplanet atmospheres. In most instances, molecular diffusion plays a dominant role, while photochemical processes also contribute significantly in certain cases.

\subsection{$\mathrm{H_2O}$}

\begin{enumerate}
    \item HAT-P-11 b: Our study does not corroborate the recent detection of $\mathrm{H_2O}$ \citep{basilicata2024gaps} on this planet due to the lower abundance of $\mathrm{H_2O}$ in the atmospheric profile. This discrepancy arises from the adoption of a higher C/O ratio in our chemistry simulations. 
    \item HD 63433 c and TOI-270 d: Water undergoes photolysis to OH by $\mathrm{H_2O}$ $\rightarrow$ H + OH, and the recombination happens by OH + $\mathrm{H_2}$ $\rightarrow$ $\mathrm{H_2O}$ + H.
    \item K2-25 b:  It forms sulphuric acid by reacting with $\mathrm{SO_3}$: $\mathrm{H_2O}$ + $\mathrm{SO_3}$ $\rightarrow$ $\mathrm{H_2SO_4}$. The formation of $\mathrm{H_2O}$ comes from the OH + $\mathrm{H_2}$ $\rightarrow$ $\mathrm{H_2O}$ + H reaction. The former reaction has a larger time scale and hence, slower than the latter reaction. This led to an enhancement in $\mathrm{H_2O}$ mixing ratios in the atmosphere.
\end{enumerate}

\subsection{$\mathrm{CH_4}$}

\begin{enumerate}
    \item HAT-P-11 b and HD 63433 c: $\mathrm{CH_4}$ is decomposed into $\mathrm{CH_3}$ via the reaction: H + $\mathrm{CH_4}$ $\rightarrow$ $\mathrm{CH_3}$ + $\mathrm{H_2}$. For HAT-P-11 b, the resulting $\mathrm{CH_3}$ quickly recombines through two pathways: (1) $\mathrm{CH_3}$ + $\mathrm{H_2S}$ $\rightarrow$ $\mathrm{CH_4}$ + $\mathrm{SH}$, and (2) H + $\mathrm{CH_3}$ + M $\rightarrow$ $\mathrm{CH_4}$ + M with the first pathway covering 85\% of $\mathrm{CH_4}$ formation. In the atmosphere of HD 63433 c, $\mathrm{CH_3}$ follows the pathway 2 for $\mathrm{CH_4}$ formation.
    \item K2-25 b and TOI-270 d: On K2-25 b and TOI-270 d, it undergoes the similar reactions as HAT-P-11 b and HD 63433 c for the destruction: H + $\mathrm{CH_4}$ $\rightarrow$ $\mathrm{CH_3}$ + $\mathrm{H_2}$. However, for both of them, the formation takes through a different path with a higher rate constant than the formation reaction: $\mathrm{CH_3}$ + $\mathrm{H_2S}$ $\rightarrow$ $\mathrm{CH_4}$ + SH.
\end{enumerate}

\subsection{$\mathrm{C_2H_2}$}

\begin{enumerate}
    \item HD 63433 c: $\mathrm{C_2H_3}$ is the source and sink molecule for $\mathrm{C_2H_2}$. Acetylene reduces to $\mathrm{C_2H_3}$ through a 3-body reaction: H + $\mathrm{C_2H_2}$ + M $\rightarrow$ $\mathrm{C_2H_3}$ + M and the same molecule forms acetylene by $\mathrm{C_2H_3}$ + H $\rightarrow$ $\mathrm{C_2H_2}$ + $\mathrm{H_2}$.
\end{enumerate}

\subsection{$\mathrm{CO}$}

\begin{enumerate}
 \item HD 63433 c: CO is important for the context of HD 63433 c as it is a photochemical product. It ends up forming HCO and SCO differently by 3 reactions: (1) H + CO + M $\rightarrow$ HCO + M (51\%) (2) CO + $\mathrm{CH_3S}$ $\rightarrow$ SCO + $\mathrm{CH_3}$ (28\%), and (3) S + CO + M $\rightarrow$ SCO + M (21\%). Similarly, these two molecules form CO separately either through photochemical destruction or through recombination: (1) HCO $\rightarrow$ H + CO (28\%) (2) HCO + H $\rightarrow$ CO + $\mathrm{H_2}$ (20\%), and (3) CS + SCO $\rightarrow$ SCO + $\mathrm{CS_2}$ + CO (38\%).
 \item K2-25 b and TOI-270 d: For both planets, CO forms SCO by following the pathway 2 from HD 63433 c: CO + $\mathrm{CH_3S}$ $\rightarrow$ SCO + $\mathrm{CH_3}$, On K2-25 b, SCO forms back to CO again through the pathway: SCO + SH $\rightarrow$ CO + $\mathrm{HS_2}$. On TOI-270d, two major reaction pathways form CO: (1) SCO + SH $\rightarrow$ CO + $\mathrm{HS_2}$ (47\%) and (2) SCO $\rightarrow$ S + CO (38\%) (photochemical pathway).
\end{enumerate}

\subsection{$\mathrm{CO_2}$}

\begin{enumerate}
\item Hd 63433 c: $\mathrm{CO_2}$ follows a usual chemical pathway and a photodissociation process to form CO: (1) $\mathrm{CO_2}$ + $\mathrm{CH_2}$ $\rightarrow$ $\mathrm{H_2CO}$ + CO (64\%) and (2) $\mathrm{CO_2}$ $\rightarrow$ CO + O (36\%). CO forms back to $\mathrm{CO_2}$ following: $\mathrm{CO}$ + OH $\rightarrow$ H + $\mathrm{CO_2}$.
 \item TOI-270 d: $\mathrm{CO_2}$ undergoes $\mathrm{CO_2}$ + CH $\rightarrow$ HCO + CO as the major destruction process. CO reacts with OH to form $\mathrm{CO_2}$: $\mathrm{CO}$ + OH $\rightarrow$ H + $\mathrm{CO_2}$.
\end{enumerate}

\subsection{$\mathrm{HCN}$}

\begin{enumerate}
 \item HD 63433 c: HCN forms $\mathrm{H_2CN}$ by a 3-body reaction: HCN + H + M $\rightarrow$ $\mathrm{H_2CN}$ + M which later recombines to HCN by: $\mathrm{H_2CN}$ + H $\rightarrow$ HCN + $\mathrm{H_2}$ (75\%). There is another reaction that produces HCN from CN by: $\mathrm{H_2}$ + CN $\rightarrow$ HCN + $\mathrm{H}$ (25\%).
\end{enumerate}

\subsection{$\mathrm{NH_3}$}

\begin{enumerate}
 \item HAT-P-11 b and K2-25 b: On HAT-P-11 b, $\mathrm{NH_3}$ has a long lifetime to its destruction processes, which form $\mathrm{NH_2}$ via the reactions: (1) SH + $\mathrm{NH_3}$ $\rightarrow$ $\mathrm{H_2S}$ + $\mathrm{NH_2}$ (62\%) and the photochemical reaction: (2) $\mathrm{NH_3}$ $\rightarrow$ $\mathrm{NH_2}$ + H (38\%). For, K2-25 b, the same pathways are followed for the destruction of $\mathrm{NH_3}$ with 73\% contribution from pathway 1 and 27\% contribution from pathway 2. On both planets, $\mathrm{NH_2}$ efficiently recombines to form $\mathrm{NH_3}$ through the reaction: $\mathrm{NH_2}$ + $\mathrm{H_2}$ $\rightarrow$ $\mathrm{NH_3}$ + H.
 \item HD 63433 c and TOI-270 d: Unlike HAT-P-11 b and K2-25 b, $\mathrm{NH_3}$ on these planets follows a single pathway to form $\mathrm{NH_2}$, specifically through pathway 2: $\mathrm{NH_3}$ $\rightarrow$ $\mathrm{NH_2}$ + H. However, $\mathrm{NH_2}$ undergoes a similar recombination process to revert back to $\mathrm{NH_3}$: $\mathrm{NH_2}$ + $\mathrm{H_2}$ $\rightarrow$ $\mathrm{NH_3}$ + H.
\end{enumerate}

\subsection{$\mathrm{CS_2}$}

\begin{enumerate}
 \item TOI-270 d: $\mathrm{CS_2}$ forms CS through photo dissociation: $\mathrm{CS_2}$ $\rightarrow$ CS + S. CS recombines with SCO to form $\mathrm{CS_2}$: CS+ SCO $\rightarrow$ $\mathrm{CS_2}$ + CO.
\end{enumerate}

\subsection{$\mathrm{H_2S}$}

\begin{enumerate}
    \item HAT-P-11 b and HD 63433 c: For both planets, $\mathrm{H_2S}$ forms SH by: $\mathrm{H_2S}$ + H $\rightarrow$ $\mathrm{H_2}$ + SH. On HAT-P-11 b, it gets recombined to $\mathrm{H_2S}$ by: (1) $\mathrm{SH}$ + SH $\rightarrow$ S + $\mathrm{H_2S}$ (41\%) and (2) H + $\mathrm{SH}$ + M $\rightarrow$ $\mathrm{H_2S}$ + M (59\%). On HD 63433 c, the recombination follows the pathway 2: H + $\mathrm{SH}$ + M $\rightarrow$ $\mathrm{H_2S}$ + M.
    \item K2-25 b and TOI-270 d: On K2-25 b and TOI-270 d, $\mathrm{H_2S}$ follows multiple pathways to destroy itself. On TOI-270 d, it covers a total of three reactions: (1) $\mathrm{H_2S}$ + H $\rightarrow$ $\mathrm{H_2}$ + SH (60\%), (2) $\mathrm{H_2S}$ $\rightarrow$ H + SH (27\%), and (3) $\mathrm{H_2S}$ + S $\rightarrow$ $\mathrm{S_2}$ + $\mathrm{H_2}$ (13\%). On K2-25 b, it follows reaction 1 with 81\% contribution and reaction 3 with 19\% contribution. However, on both planets, $\mathrm{H_2S}$ formation is solely led by SH through the following pathway: SH + SH $\rightarrow$ S + $\mathrm{H_2S}$.
\end{enumerate}


\end{document}